\documentclass[12pt]{article}
\usepackage{a4}
\usepackage{epsfig}
\def\Journal#1#2#3#4{{#1} {\bf #2}, #3 (#4)}

\def\PRL{\em Phys. Rev. Lett.}
\def\PRD{{\em Phys. Rev.} D}
\def\PRP{{\em Phys. Reports}}

\begin{document}
\newcommand{\newc}{\newcommand}
\newc{\R}{$R$}
\newc{\charginom}{M_{\tilde \chi}^{+}}
\newc{\mue}{\mu_{\tilde{e}_{iL}}}
\newc{\mud}{\mu_{\tilde{d}_{jL}}}
\newc{\barr}{\begin{eqnarray}}
\newc{\earr}{\end{eqnarray}}
\newc{\beq}{\begin{equation}}
\newc{\eeq}{\end{equation}}
\newc{\ra}{\rightarrow}
\newc{\lam}{\lambda}
\newc{\eps}{\epsilon}
\newc{\gev}{\,GeV}
\newc{\tev}{\,TeV}
\newc{\eq}[1]{(\ref{eq:#1})}
\newc{\eqs}[2]{(\ref{eq:#1},\ref{eq:#2})}
\newc{\etal}{{\it et al.}\ }
\newc{\eg}{{\it e.g.}\ }
\newc{\ie}{{\it i.e.}\ } 
\newc{\nonum}{\nonumber}
\newc{\lab}[1]{\label{eq:#1}}
\newc{\dpr}[2]{({#1}\cdot{#2})}
\newc{\gsim}{\stackrel{>}{\sim}}
\newc{\lsim}{\stackrel{<}{\sim}}
\begin{titlepage}
\voffset=-20pt
\begin{figure}[h]
\includegraphics*[scale=1.,angle=0,bb=222 438 402 552 ]
{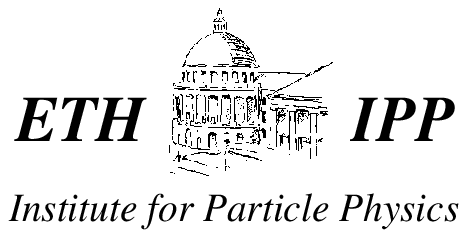}
\end{figure}
\begin{picture}(100,-100)(10,20)
\put (350.,130.){\bf ETHZ-IPP  PR-99-06}
\end{picture}
\begin{picture}(100,-100)(10,20)
\put (245.,112.){\bf July 19, 1999}
\end{picture}

\begin{center}
{\bf \LARGE Status and Prospects of Supersymmetry Searches at Colliders} 
\end{center}
\bigskip

\begin{center}
{\large Michael Dittmar}
\end{center}
\bigskip

\begin{center}
Institute for Particle Physics (IPP), ETH Z\"{u}rich, \\
CH-8093 Z\"{u}rich, Switzerland
\end{center}

\begin{abstract}
\noindent 
Experiments at particle colliders 
have reached center of mass energies well above 100 GeV, equivalent 
to temperatures which existed shortly after the big bang.
These experiments, testing the initial conditions of the 
universe have, with great precision, established  
the Standard Model of Particle Physics. In contrast,
the existence of the Higgs boson and perhaps Supersymmetry 
remain speculative, as todays searches have failed to find 
signs of their existence.
However, the next generation of high energy collider experiments and 
especially CERN's 14 TeV LHC, expected to start operation in the year 2005, 
should lead either to the discovery of the Higgs and
Supersymmetry or disprove todays theoretical ideas. 
\end{abstract}
\bigskip

\begin{center}
{\it \large Lecture given at the \\
6~$^{eme}$ Colloque Cosmologie \\
Observatoire de Paris \\
Paris, June 16--18 1999}
\end{center}
\end{titlepage}

\section{Introduction}

Experiments at particle colliders have established,
with great precision, the validity  
of todays Standard Model (SM) of particle physics as an accurate description 
of high energy physics up to mass scales of a few 100 GeV.
Among the most precise tests~\cite{ew99tamp}, one finds the $W$ and $Z$ boson 
masses and their coupling strength to fermions. 
Combining all these measurements and staying within the SM 
one starts being sensitive to the last missing SM particle, the
Higgs boson. Furthermore, accurate measurements of the $Z$ boson 
width, allowed to determine the number of light neutrino families 
to be 2.9835$\pm$0.0083 and to establish 
lepton universality with an accuracy well below the 1\% level. 
These precise laboratory measurements 
lead thus to the requirement that a more complete model
has to include todays SM as a low energy approximation.
It is often said that physics beyond the 
SM is required because 
some fundamental questions are not addressed by the 
SM. These problems are related to the 
``unnatural'' mass splitting between the known 
fundamental fermions with neutrino masses close to 0 eV 
and $\approx$ 175$\cdot 10^{9}$ eV for the top quark and  
the so called hierarchy problem or fine tuning problem of the Standard Model. 
The hierarchy problem originates from theoretical ideas to 
extrapolate todays knowledge at mass scales of 
a few 100 GeV to energy scales of about $10^{15}$ GeV and more.  
A purely theoretical approach to this extrapolation has lead theorists 
to Supersymmetry and the so--called 
``Minimal Supersymmetric Standard Model'' (MSSM)~\cite{nilles},  
which could solve some of these conceptual problems by
introducing super-symmetric partners to every known 
boson and fermion and at least an additional Higgs multiplett. 
As these SUSY particles should have been produced abundantly
shortly after the Big Bang, Supersymmetry with R--parity 
conservation offers a lightest stable super symmetric particle
as the ``cold dark matter'' candidate.

Despite the variety of SUSY models with largely unconstrained masses of 
SUSY particles and the absence of any indications for Supersymmetry, 
searches for Supersymmetry at existing and sensitivity estimates 
of future collider experiments became an important aspect 
of high energy physics. 

This report is structured as follows: its starts with an overview
of experimentation at high energy colliders, which includes some 
recent experimental highlights, we discuss basics concepts of 
Supersymmetry and the applied search strategies.    
We than describe a few examples of 
experiments with negative results at LEP II and the TEVATRON 
and give an outlook to future perspectives at the LHC.
  
\section {Experimentation and Experiments at high 
Energy Colliders}

The high energy frontier of particle colliders are currently covered 
by the $e^{+}e^{-}$ collider LEP at CERN and 
the proton--antiproton collider TEVATRON 
at FERMILAB. In contrast to 
$e^{+}e^{-}$ colliders which investigate nature directly
at the available center--of--mass energy $\sqrt{s}$, 
proton colliders study of quark and gluon 
collisions over a wide $\sqrt{s_{eff}}$ range. The
maximal effective $\sqrt{s_{eff}}$ is however, 
depending on the available luminosity, 
about a factor of $\approx$4--6 smaller than
the nominal $\sqrt{s}$ of hadron--hadron collisions.

The LEP collider is currently running at
center--of--mass energies, $\sqrt{s}$, of 196 GeV and will soon reach 
$\sqrt{s} \approx$ 200 GeV. The TEVATRON experiments CDF and D0
have collected data corresponding to a luminosity of 
about 100 pb$^{-1}$ per experiment at center--of--mass energies 
of 1.8 TeV. The upgraded TEVATRON with its improved experiments is 
expected to restart running in the year 
2000 at 2 TeV center--of--mass energies and should provide 
luminosity of 1--2 fb$^{-1}$ per year and experiment.
Around the year 2005 the LHC, CERN's large 
hadron collider, is expected to come into operation. 
The LHC is a proton--proton collider with 14 TeV center--of--mass energies 
and high luminosity. The LHC will allow to increase the sensitivity to 
new physics well into the TeV mass range.     
\begin{figure}[htb]
\begin{center}
\includegraphics*[scale=0.3,angle=0,bb=00 100 600 600 ]
{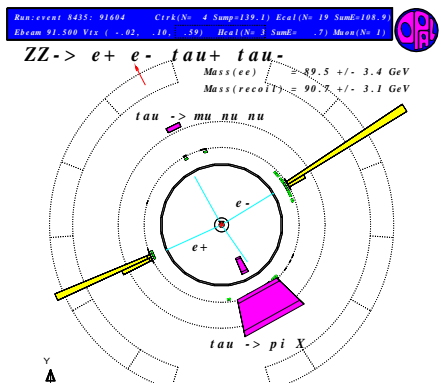}
\caption[fig1]{Example of the reaction
$e^{+}e^{-} \rightarrow Z^{0}Z^{0} \rightarrow e^{+}e^{-} \tau^{+} \tau^{-}$
from OPAL at LEP~\cite{opalevent}. 
The mass of the $e^{+}e^{-}$ pair is accurately 
measured.  The mass of the $\tau\tau$ system is determined indirectly using
the recoil mass, estimated from the known $\sqrt{s}$ and 
requiring energy and momentum conservation. 
}
\end{center}
\end{figure}
%\newpage
\begin{figure}[htb]
\begin{center}
\includegraphics*[scale=0.25,angle=0,bb=00 00 600 600 ]
{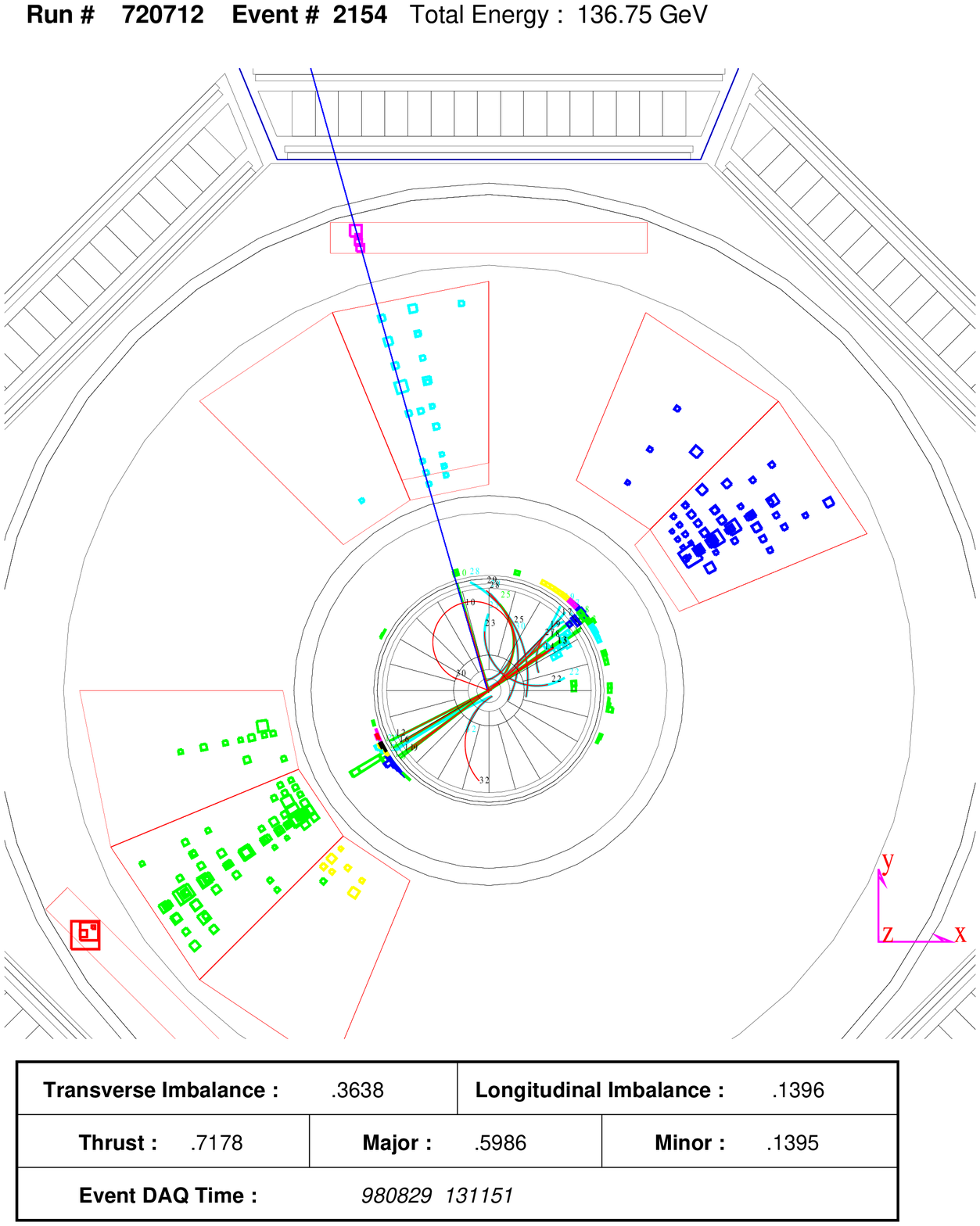}
\caption[fig2]{Example of the reaction
$e^{+}e^{-} \rightarrow W^{+}W^{-} \rightarrow q \bar{q} \mu^{-} \nu$
from L3 at LEP~\cite{l3event}. 
One sees an isolated stiff track, identified as 
a muon and two almost back to back hadron jets assigned to the 
hadronic $W$ decay. The measured event indicates further that something 
invisible, the neutrino, is recoiling against the reconstructed muon. 
}
\end{center}
\end{figure}
 
Modern Collider experiments have essentially a cylindrical structure with 
an outer radius of 5--7 m and a length between 10--25 m. The onion like 
structure of these experiments consists of:  
\begin{enumerate}
\item
Precision detectors which measure precisely 
the trajectories of charged particles. These detectors are 
embedded in a magnetic field with a z--axis along the beam direction,
which allows to measure the momentum of charged particles.  
\item electromagnetic and hadron calorimeters which measure accurately the 
impact position and the energy electrons, photons and charged or 
neutral hadrons and,
\item 
muon detection systems surround the calorimeters and hadron absorbtion length, 
which measure the position and direction  
of muons.
\end{enumerate}  
While todays collider experiments are constructed and operated   
by international collaborations of up to 500 physicists one 
expects that tomorrows LHC experiments will unite nearly 2000 physicists. 

%\newpage
Depending only slightly on the main aim of the experiments,  
the detectors allow to measure the energy and momentum of 
long lived charged 
($\pi^{\pm}, K^{\pm}, p, \bar{p}, e^{\pm}$ and $\mu^{\pm}$)  
as well as the neutral particles $\gamma, K^{0}$ and $n, \bar{n}$.
These individual particles can than be combined to search for mass 
peaks of short lived particles, $\tau$ decay products and bunches of hadrons 
identified as jets. These jets can be separated 
using their characteristic lifetime and kinematics into 
light quark (u,d, and s) or gluon jets, c(harm) and b(eauty) flavoured 
jets. Furthermore, modern experiments have achieved essentially a 
$4\pi$ angular coverage for the various individual energy 
and momentum measurements, which allows the determination of    
the missing energy and momentum due to ``invisible'' neutrinos or 
neutrino like objects.
Examples of a few interesting 
events, produced at LEP II and the TEVATRON, are 
shown in Figures 1--4. These events indicate how the 
original physics process can be reconstructed from the detectable particles.

\begin{figure}[htb]
\begin{center}
\includegraphics*[scale=0.35,angle=0,bb=100 200 500 600 ]
{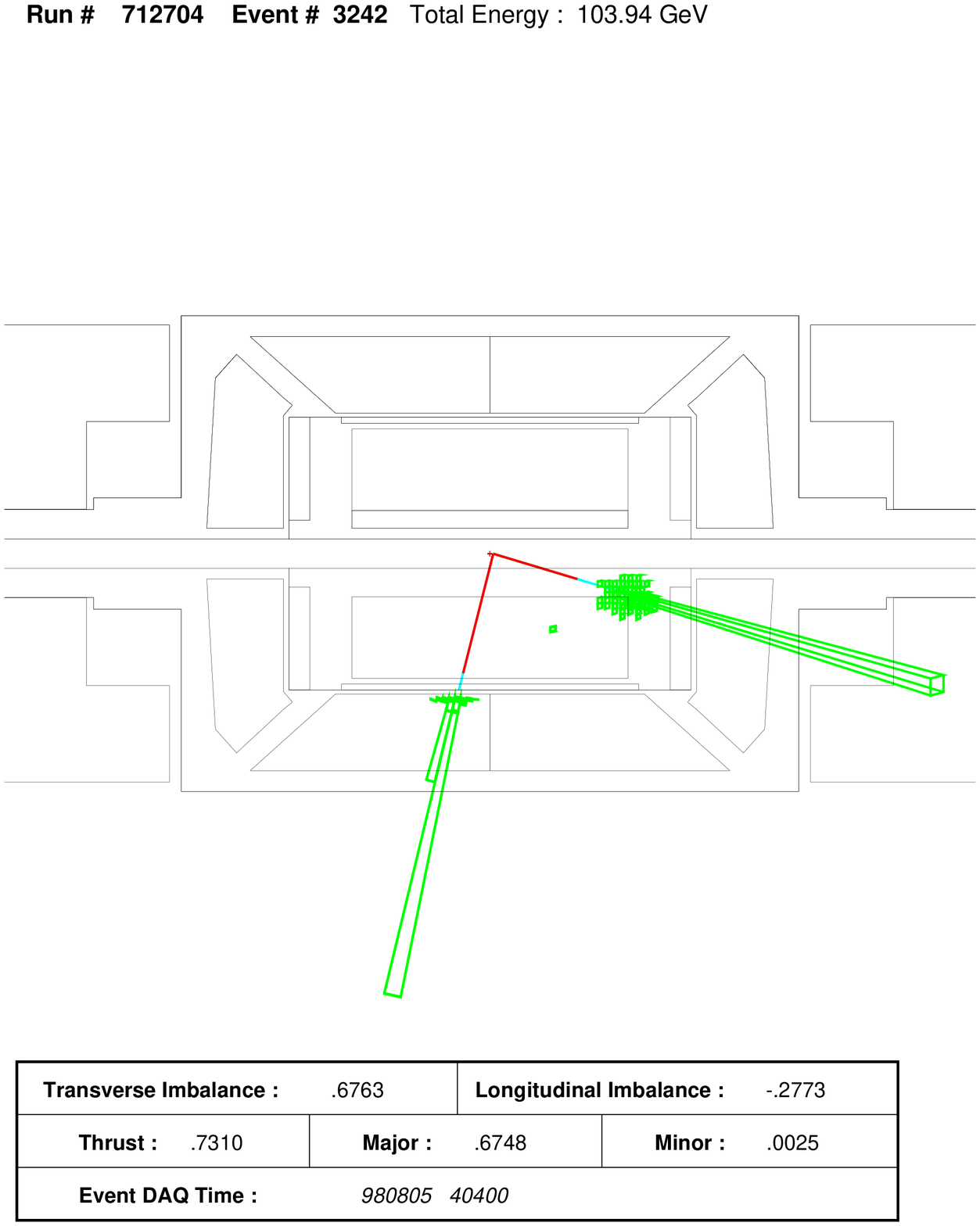}
\includegraphics*[scale=0.2,angle=0,bb=0 50 600 660 ]
{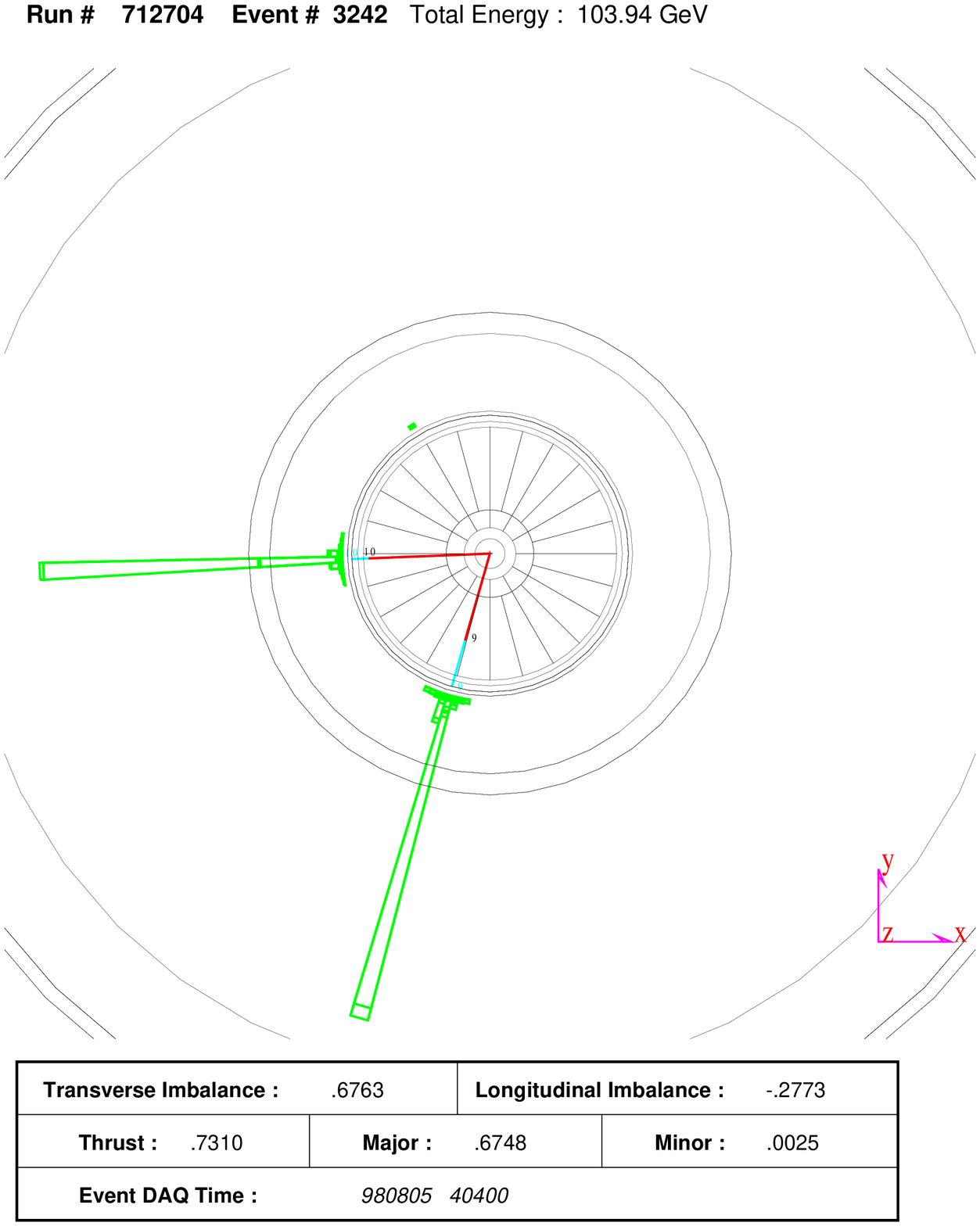}
\caption[fig3]{Example from L3 at LEP~\cite{l3event} of the reaction
$e^{+}e^{-} \rightarrow W^{+}W^{-} \rightarrow e^{+} \nu e^{-} \nu$
in a plane including the 
beam axis and in the plane transverse to the beam.
The observed event kinematics indicates clearly that ``something'' 
must escape undetected.
}
\end{center}
\end{figure}

\begin{figure}[htb]
\begin{center}
\includegraphics*[scale=0.25,angle=0,bb=00 50 600 600 ]
{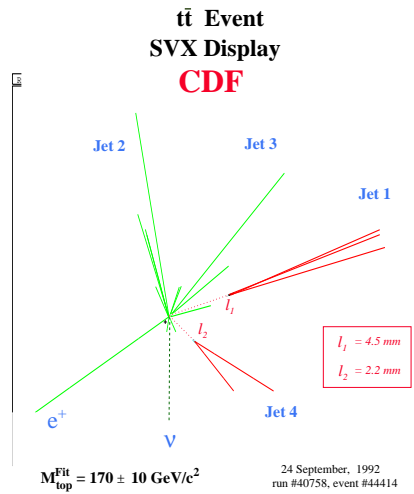}
\caption[fig4]{
Example of the reaction
$p \bar{p} \rightarrow t \bar{t} \rightarrow e^{+} \nu b q\bar{q} b$
as seen at the TEVATRON with the CDF detector~\cite{cdfevent}.
Reconstructed charged particle trajectories are shown in 
an enlarged view of the interaction region. Two separated 
vertices, assigned to the decay products of b--jets, are visible. 
}
\end{center}
\end{figure}
%\newpage
\clearpage 
\subsection{Highlights from Collider Experiments}
Despite the non observation of physics beyond the 
SM, recent experiments at particle colliders 
gave a large variety of impressive results.
Especially remarkable are the measurements of the $Z$ boson 
parameters with the resulting 
number of light neutrino families being 2.9835$\pm$0.0083, 
the discovery of the top quark at the TEVATRON,  
the measured cross section of the reaction $e^{+}e^{-} \rightarrow WW$
at LEP II and the observed energy dependence of the strong coupling 
constant $\alpha_{s}$.
Some experimental results and the corresponding theoretical 
expectations are shown in Figures 5--7.
\begin{figure}[htb]
\begin{center}
\includegraphics*[scale=0.4,angle=-90,bb=00 00 800 900 ]
{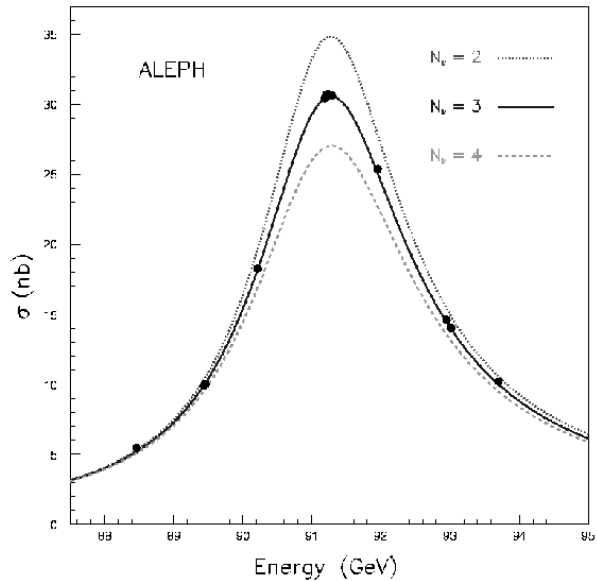}
\caption[fig5]{
Measured $Z$ boson cross section for different 
$e^{+}e^{-}$ center--of--mass energies (from ALEPH) and 
the theoretical curves assuming 2,3 and 4 light neutrinos~\cite{alephnus}.   
}
\end{center}
\end{figure}
  
\begin{figure}[htb]
\begin{center}
\epsfig{file=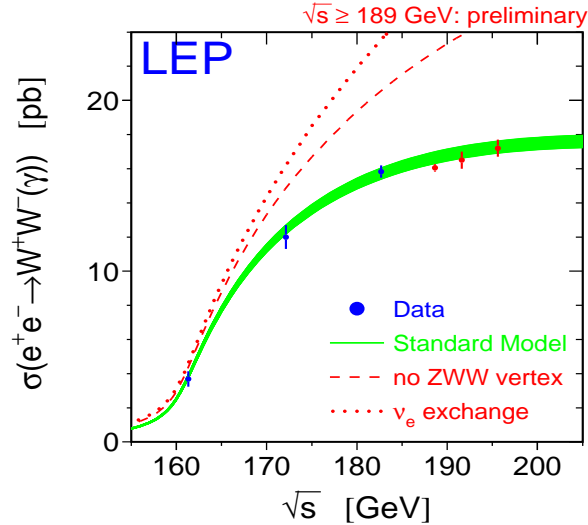,
width=8.cm,height=8.cm}
\caption[fig6]{Combination of the measured cross sections for the 
reaction $e^{+}e^{-} \rightarrow WW$ from the four LEP experiments
(ALEPH, DELPHI, L3 and OPAL) together with the SM expectation 
and some excluded alternatives~\cite{ew99tamp}. 
}
\end{center}
\end{figure}
 
\begin{figure}[htb]
\begin{center}
\epsfig{file=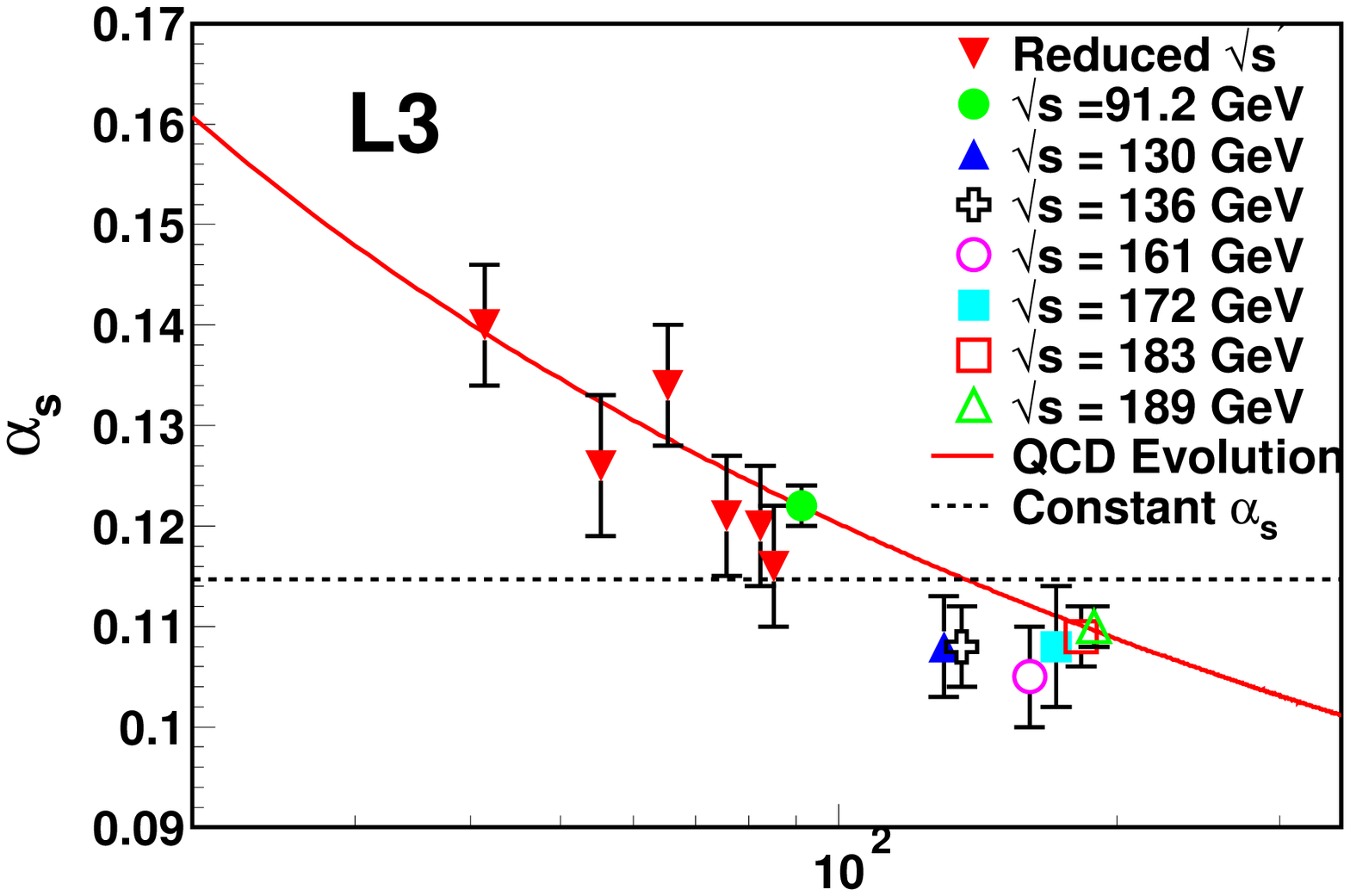,
width=14.cm,height=8.cm}
\caption[fig7]{
An example of the measured and expected energy dependence of 
$\alpha_{s}$ from L3 at LEP~\cite{l3alphas}. 
}
\end{center}
\end{figure}
 
Combining the results of the large variety of collider measurements, 
one is forced to accept that the SM is at least an 
excellent approximation of nature. Furthermore, 
one starts, as shown in Figure 8, to have 
some indirect constraints on the Higgs boson mass.
These indirect constraints are from the combination of 
the different measurements of electroweak observables, 
like the various asymmetry measurements in $Z$ decays and
the masses of the $W^{\pm}$ and the top quark.
The accuracy of this procedure is however limited as there is only a soft 
logarithmic Higgs mass dependence. 
Nevertheless, assuming that the Higgs mass is the only unknown
SM parameter, a fit to all precision data 
constrains the Higgs mass to  
$92\pm^{78}_{45}$ GeV and with a confidence level 
of 95\% c.l. to less than about 
245 GeV~\cite{ew99tamp}. This result agrees with 
Higgs mass estimates of $\approx 160\pm 20$ GeV~\cite{mhiggssm2}, 
which assume the validity of the 
SM up to very large mass scales like the Planck scale.
It agrees also with 
expectations from Supersymmetry with the minimal Higgs sector 
where the lightest Higgs must have a mass of less than $\approx$ 130 GeV.
 
\begin{figure}[htb]
\begin{center}
\mbox{
\epsfig{file=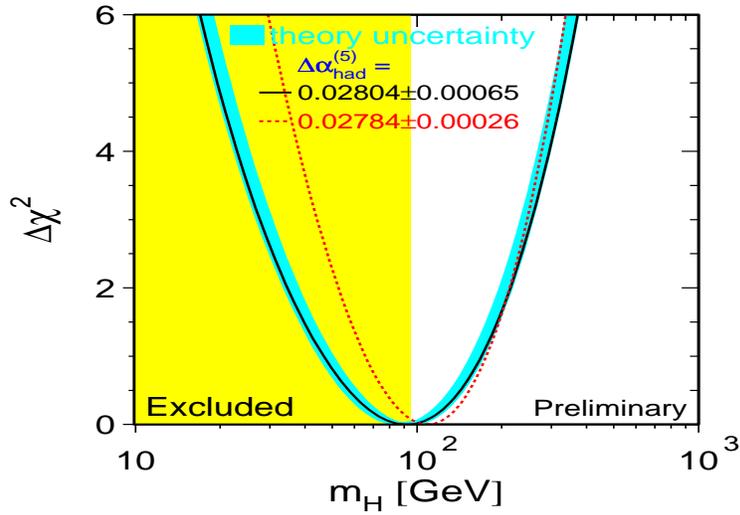,
height=8 cm,width=10cm}
}
\end{center}
\caption[fig8]{$\Delta \chi^{2}$ result of 
a SM fit to all electroweak observables assuming to have the 
Higgs mass as the only remaining free parameter~\cite{ew99tamp}.
The 95\% c.l. SM Higgs mass upper limit is 
obtained from a $\chi^{2}$ variation 
of 4 with respect to the minimum which corresponds to a value of 22 for 
15 degrees of freedom.}
\end{figure}

The precise measurements of the energy dependence of the 
strong, the electromagnetic and the weak coupling parameters 
indicate comparable couplings at energies close to $10^{15}$ GeV.
However, the expectation from the simplest Grand-Unification 
theories of a perfect matching is now excluded. It might however 
be achieved if some new physics, like Supersymmetry, 
exists at nearby mass scales. Another indication of
physics beyond todays SM comes from the 
observed ``unnatural'' large mass splitting between the otherwise 
identical fermion families which cover at least 11 orders of magnitude 
and the ``large'' number of free parameters within the SM and 
the exclusion of gravity.     
\newpage
\clearpage
\section {Beyond the Standard Model of Particle Physics:
Supersymmetry, SUSY Models and SUSY Signatures}

Among the possible extensions of the Standard Model the 
Minimal Supersymmetric Standard Model (MSSM)~\cite{nilles} 
is usually considered 
to give the most serious theoretical frame. The attractive features 
of this approach are:
\begin{itemize}
\item It is quite close to the existing Standard Model. 
\item It explains the so called hierarchy problem of the Standard Model.
\item It allows to calculate.   
\item Predicts many new particles and thus ``Nobel Prizes'' for the 
masses.
\end{itemize}

An example for  
the small difference between the SM and the MSSM in terms of 
electroweak observables is shown in 
Figure 9~\cite{hollik98}, which compares the measurements of the
$W$--boson and the top--quark masses with predictions of
the SM and the MSSM. 
\begin{figure}[htb]
\begin{center}
\includegraphics*[scale=0.65,bb=0 150 600 650 ]
{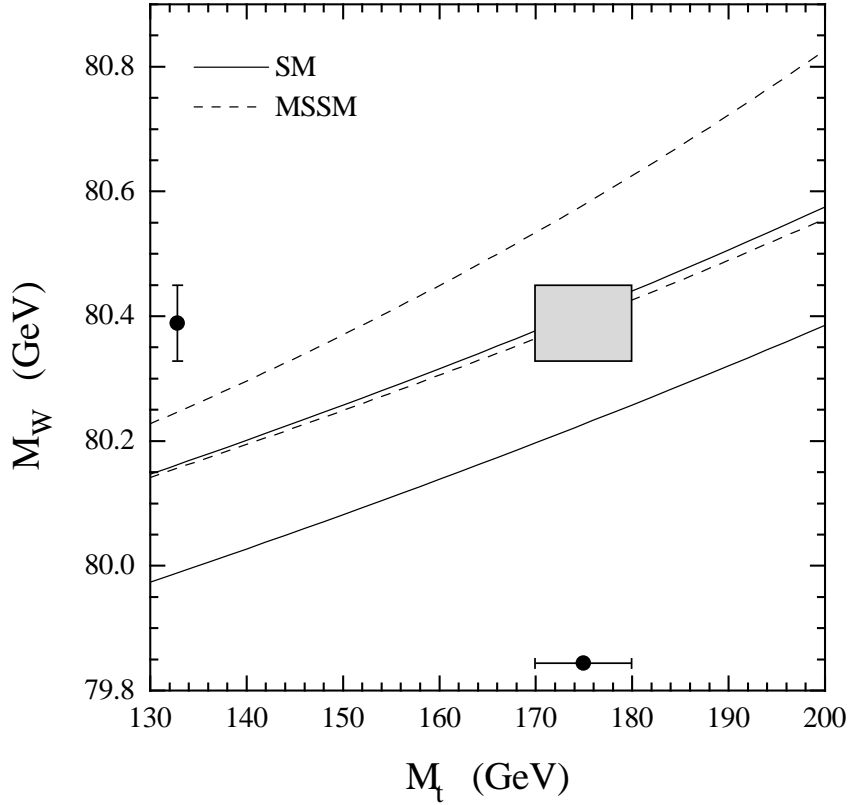}
\end{center}
\caption[fig9]{Expected relation between 
$M_{W}$ and $M_{top}$ in the 
Standard Model and  
the MSSM, the bounds are from the non--observation of Higgs or SUSY 
particles at LEP II~\cite{hollik98}.
The 1998 experimental area, with 
$M_{W}=80.39 \pm 0.06$ GeV and $M_{top}=174 \pm 5$ GeV
is also indicated.
}
\end{figure}

Unfortunately todays data,  
$M_{W}=80.394 \pm 0.042$ GeV and $M_{top}=174 \pm 5$ GeV~\cite{ew99tamp},   
favor an area which is perfectly consistent with both models.
Similar conclusion can be drawn from other comparisons of 
todays precision measurements with  
the SM and the MSSM.

A large number of new heavy particles should exist within the MSSM model.
In detail, one expects spin 0 partners, called sleptons and squarks 
for every quark and lepton  
and spin 1/2 partners, called gluinos, charginos and neutralinos, for  
the known spin 1 bosons and for the hypothetical scalar Higgs bosons.
Due to identical quantum numbers, some mixing between the different 
neutralinos and charginos might exist. 
In addition, at least 5 Higgs bosons ($h^{0}, H^{0}, A^{0}$ and $H^{\pm}$) 
are required. The masses of these Higgs bosons are strongly related. 
Essentially one needs to know ``only'' the mass of $h^{0}$ and one other  
Higgs boson or the mass of one Higgs boson and $\tan \beta$, 
the ratio of the higgs vacuum expectation values. 
 
Experimental searches for Supersymmetry can thus be divided into 
a) the MSSM Higgs sector and b) the direct SUSY particle search.

The advantages of searches for a Higgs boson are 
that at least one Higgs boson with a mass smaller than about 
130 GeV~\cite{lowmassh} 
should exist and that cross sections and 
decay modes can be calculated accurately as a function of the mass 
and $\tan \beta$. The disadvantage is however that,
in order to distinguish between the SM and Supersymmetry,
at least two MSSM Higgs bosons need to be discovered.

In contrast to Higgs searches, searches for SUSY 
particles can look for a variety of SUSY particles and 
the discovery of one SUSY particle could be a proof of Supersymmetry.
Unfortunately the masses of SUSY particles  
cannot be predicted and values far beyond todays and perhaps even tomorrows 
center--of--mass energies are possible. Furthermore, 
having over 100 free SUSY parameters 
a large variety of SUSY signatures needs to be studied.    

As will be discussed in section 4.1, todays negative search results for Higgs
particles at LEP II indicate that the mass of 
the lightest SUSY Higgs must be greater than about 80-90 GeV. 
The absence of any indication for 
super symmetric particles at LEP II and the TEVATRON, discussed below
imply mass limits of about 90 GeV for sleptons, 95 GeV for 
charginos and about 200 GeV for squarks and gluinos~\cite{treille}.

\subsection {Signatures of SUSY Particles}

Essentially all signatures related to the MSSM are
based on the consequences of R--parity conservation~\cite{nonrparity}. 
R--parity is a 
multiplicative quantum number like ordinary parity. The R--parity of 
the known SM
particles is 1, while the one for the SUSY partners is --1. 
As a consequence, SUSY particles have to be produced in pairs.
Unstable SUSY particles decay, either directly or via 
some cascades, to SM particles and the lightest super symmetric
particle, LSP, required by cosmological arguments to be 
neutral. Such a massive LSP's, should have been abundantly produced 
after the Big Bang and is currently considered 
to be ``the cold dark matter'' candidate. 

This LSP, usually assumed to be the lightest neutralino 
$\tilde{\chi}^{0}_{1}$ has neutrino like interaction cross sections
and cannot be observed in collider experiments. 
Events with a large amount of missing energy and momentum are 
thus the SUSY signature in collider experiments.
Due to neutrinos produced in weak decays 
of for example $\tau$ leptons and measurement errors, 
the missing energy and momentum signature alone are 
usually not sufficient to identify SUSY particles. 

However, SM backgrounds can be strongly reduced 
if the decay kinematics of heavy particles are exploited. 
The decay products of heavy particles obtain a relatively large 
$p_{\perp}$ with respect to the momentum vector of the 
decaying particle and can thus be emitted with large angles.
Consequently, the observable decay products 
of pair produced heavy SUSY particles should thus 
be seen in non back--to--back events.
Due to the detection hole close to the beam line,
missing momentum along the beam direction is also be expected  
for standard physics reactions. SUSY searches 
concentrate thus on the missing momentum 
in the plane transverse to the beam direction and require 
usually some non back--to--back signature in this x--y plane.
Essentially all the characteristics of SUSY searches, large missing 
momentum, energy and the non back--to--back signature are also used 
to select 
$e^{+}e^{-} \rightarrow WW \rightarrow \ell^{+} \nu \ell^{-} \bar{\nu}$ events 
at LEP II as visualized in events like the one shown in Figure 3.
Due to the relatively large $W$ mass and cross section, SUSY searches 
at LEP II and other colliders need to consider especially the 
potential backgrounds from the SM reaction $e^{+}e^{-} \rightarrow WW$.

Possible SUSY search examples, which exploit the above signatures 
are the pair production of sleptons with their subsequent decays, 
$e^{+}e^{-} \rightarrow \tilde{\ell}^{+}\tilde{\ell}^{-}$
and $ \tilde{\ell} \rightarrow \ell \tilde{\chi}^{0}_{1}$.
Such events would appear as events with a pair of isolated electrons or muons 
with high $p_{t}$ and large missing transverse energy.

Starting from the MSSM, the so called minimal model, one counts 
more than hundred free parameters. So many unconstrained 
parameters do not offer a good guidance for experimentalist
which prefer to use additional assumptions.  
The perhaps simplest approach is the MSUGRA 
(minimal super gravity) model with only five free parameters 
($m_{0}, m_{1/2}, \tan \beta, A^{0}$ and $\mu$).
Within the MSUGRA model, the masses of 
SUSY particles are strongly related to the so called universal fermion 
and scalar masses $m_{1/2}$ and $m_{0}$. The masses of the spin 1/2 
SUSY particles are directly related to $m_{1/2}$. 
One expects approximately the following mass hierarchy:
\begin{itemize}
\item
$\tilde{\chi}^{0}_{1} \approx 1/2 m_{1/2} $
\item
$\tilde{\chi}^{0}_{2}\approx \tilde{\chi}^{\pm}_{1} \approx m_{1/2} $
\item
$\tilde{g}$ (the gluino) $\approx 3 m_{1/2} $
\end{itemize}
The masses of the spin 0 
SUSY particles are related to $m_{0}$ and $m_{1/2}$ and 
allow, for some mass splitting between the ``left'' and ``right'' handed
scalar partners of the degenerated left and right handed fermions. 
One finds the following simplified mass relations:
\begin{itemize}
\item
$m(\tilde{q})$(with q=u,d,s,c and b) $ \approx \sqrt{m_{0}^{2} + 6 m_{1/2}^{2}}$
\item
$m(\tilde{\nu}) \approx m(\tilde{\ell^{\pm}})$ (left)
$ \approx \sqrt{m_{0}^{2} + 0.52 m_{1/2}^{2}}$
\item
$m(\tilde{\ell^{\pm}})$ (right) $\approx \sqrt{m_{0}^{2} + 0.15 m_{1/2}^{2}}$
\end{itemize}
The masses of the left and right handed stop quarks ($\tilde{t}_{\ell, r}$) 
might, depending on other parameters,  
show a large mass splitting. As a result, the right handed stop quark might be 
the lightest of all squarks.   
 
Following the above mass relations and using the known SUSY couplings, 
possible SUSY decays and the related signatures can be defined.
Already with the simplest MSUGRA frame one finds   
a variety of decay chains.

For example the $\tilde{\chi}^{0}_{2}$ could decay 
to $\tilde{\chi}^{0}_{2}\rightarrow \tilde{\chi}^{0}_{1}+ X$
with $X$ being:
\begin{itemize}
\item
$X= \gamma^{*} Z^{*} \rightarrow \ell^{+}\ell^{-}$
\item
$X= h^{0} \rightarrow b \bar{b}$
\item
$X= Z  \rightarrow f \bar{f}$
\end{itemize}

Other possible $\tilde{\chi}^{0}_{2}$ decay chains are 
$\tilde{\chi}^{0}_{2}\rightarrow \tilde{\chi}^{\pm(*)}_{1}+ \ell^{\pm} \nu$ 
and 
$\tilde{\chi}^{\pm(*)}_{1} \rightarrow \tilde{\chi}^{0}_{1} \ell^{\pm} \nu$
or 
$\tilde{\chi}^{0}_{2}\rightarrow \tilde{\ell}^{\pm} \ell^{\mp}$.

Allowing for higher and higher masses, even more decay channels
might open up. It is thus not possible to define all search strategies
a priori. Furthermore, possible 
unconstrained mixing angles between neutralinos, lead to 
model dependent search strategy for squarks and gluinos.

\clearpage
\section{Where we did not discover Supersymmetry} 

\subsection {The Higgs Search at LEP II and beyond}

Experiments at LEP II and $\sqrt{s} \approx 200$ GeV
will have an excellent sensitivity to the SM Higgs with 
masses of about 100--105 GeV, 
using the process $e^{+}e^{-} \rightarrow Z^* \rightarrow Z h^{0}$.
The experimental signatures for this process 
are given by the combination 
of the various decay products of a $Z$ boson and 
two b--jets or $\tau\tau$ final states coming from the decay of the Higgs
boson. Furthermore, due to kinematic constraints the mass of the 
system recoiling against the $Z$ system can be measured with an accuracy of 
about $\pm$ 2--3 GeV and a signal should show up in the 
recoil mass spectrum. The latest LEP results, obtained using the 1998
data at $\sqrt{s}= 189$ GeV and a luminosity of 
$\approx$ 180 pb$^{-1}$/experiment, do not show any signal.
Only OPAL sees a two sigma excess of events with a recoil mass
close to the $Z$ mass as shown in Figure 10~\cite{opalhiggs98}. 
\begin{figure}[htb]
\begin{center}
\epsfig{file=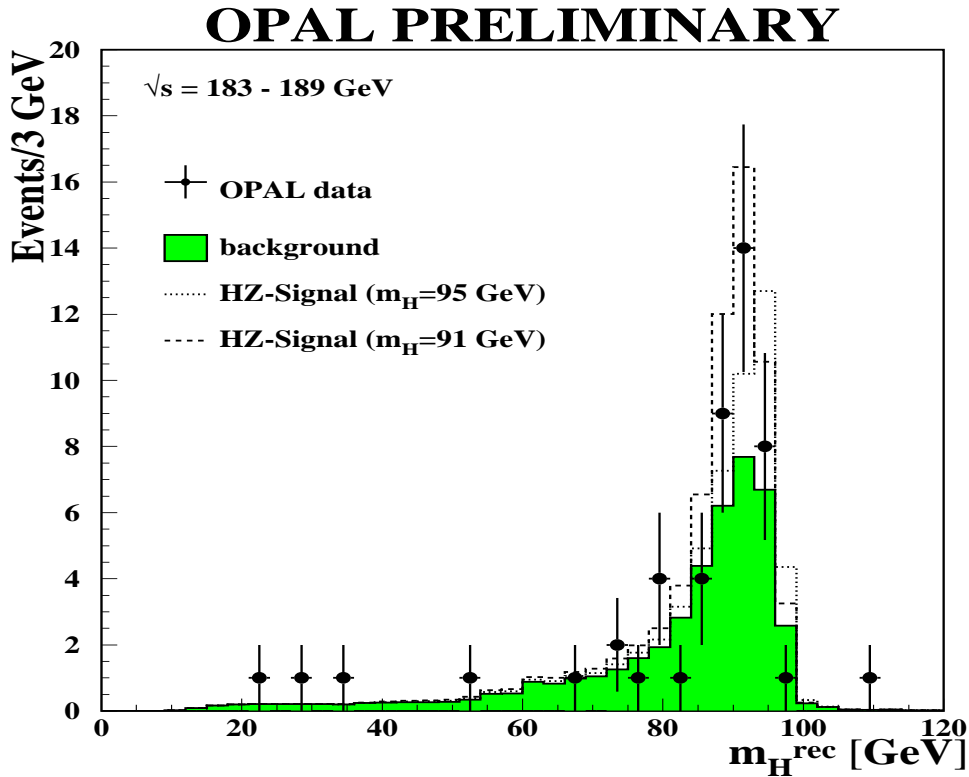,
width=14.cm,height=11cm}
\caption[fig10]
{Observed and expected mass distributions from the 
OPAL Higgs search using the data collected in the years 1997 
and 1998~\cite{opalhiggs98}.}
\end{center}
\end{figure}
The observed number of events
in the peak region is 31 compared to a background expectation of 
$\approx$ 22 events. Unfortunately the excess is neither confirmed by 
the new OPAL data collected up to July 1999~\cite{opalhiggs99} nor by 
the combination of the four LEP experiments and the 
1998 data as shown in Figure 11~\cite{martha99}. 
\begin{figure}[htb]
\begin{center}
\epsfig{file=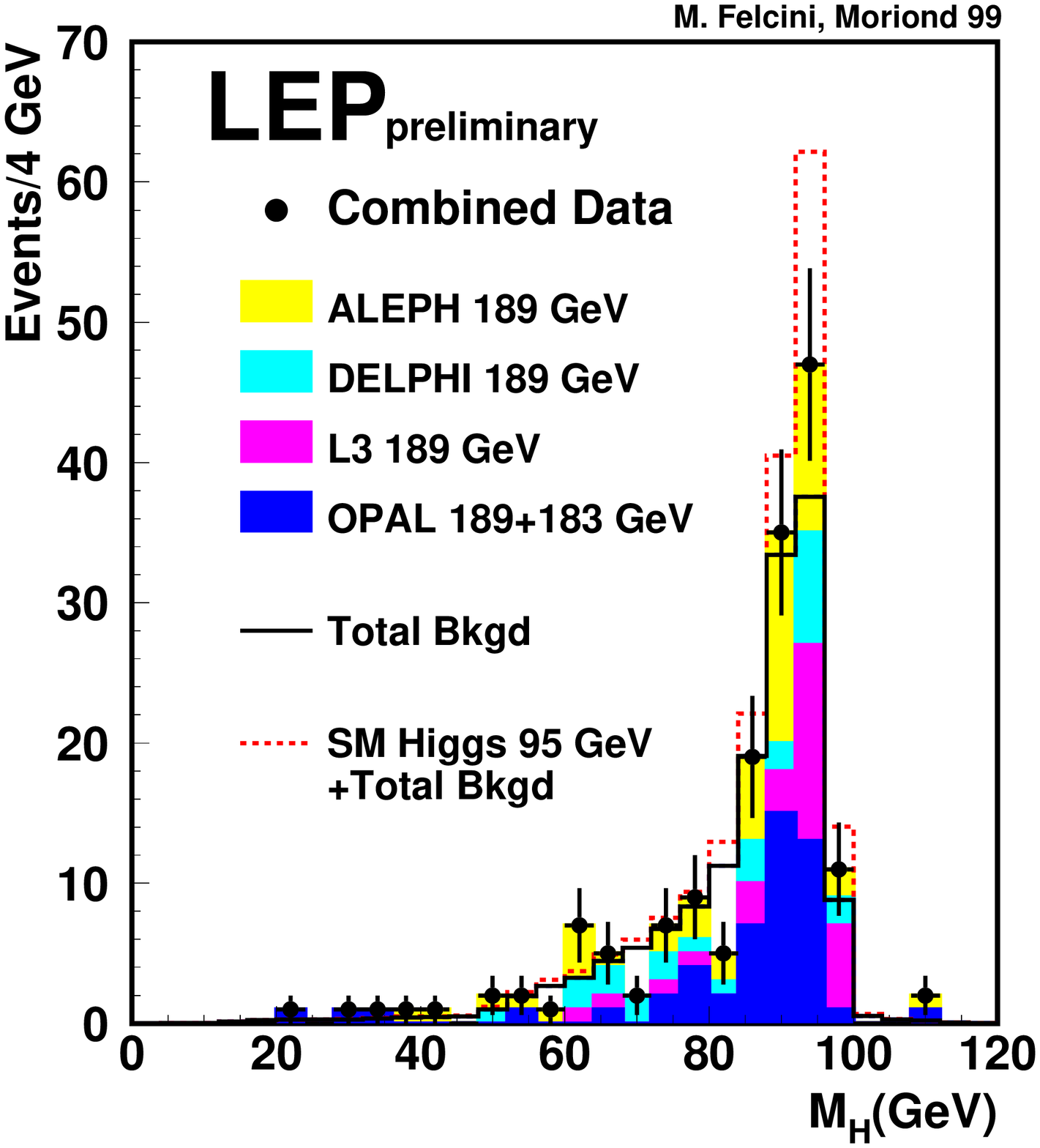,
width=14.cm,height=8cm}
\epsfig{file=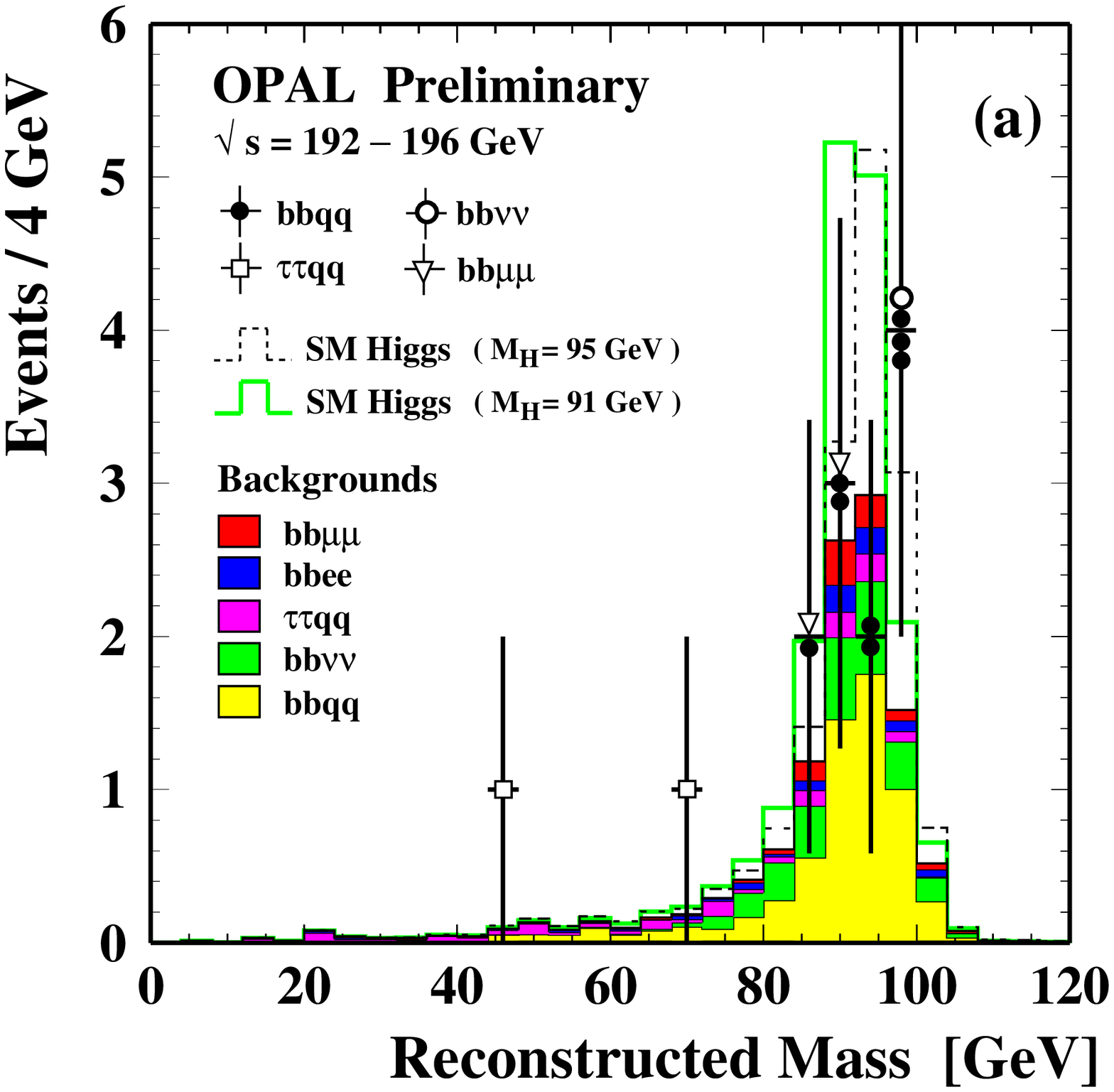,
width=13.cm,height=8cm}
\caption[fig11]
{Mass distribution for Higgs candidates from ALEPH, DELPHI, L3 
and OPAL in the 1998 data sample~\cite{martha99}  
and from the preliminary 1999 OPAL data, collected up to 
July~\cite{opalhiggs99}.
The expected mass distribution for backgrounds and a SM Higgs 
with a of 95 GeV are also shown..}
\end{center}
\end{figure}

For higher values of $\tan \beta$ the couplings of the $h^{0}$ 
to the weak bosons are reduced proportional to $\cos \beta$. 
However, the reaction 
$e^{+}e^{-} \rightarrow Z^* \rightarrow h^{0} A^{0}$, 
if kinematically allowed, appears to be detectable. This process results in   
a distinct signature of events with 
four b--jets. The search for such 4 b--jet events during the 
future LEP II running will thus give sensitivity to
masses of $M_{h}, M_{A} < 90-100$ GeV and all $\tan \beta$ values. 
The possible final SM Higgs sensitivity from the 1999/2000 
data taking at LEP II has been estimated to be 
about 105 GeV, using a luminosity of about 
$ 4 \times 200$ pb$^{-1}$ at $\sqrt{s} = 200$ GeV~\cite{hlep2f}.
One finds that this sensitivity 
translates to a Higgs sensitivity of the MSSM  
for values of $\tan \beta$ of about roughly 6--7 (3--4) 
with no (maximal) mixing
using the process $e^{+}e^{-} \rightarrow Z^* \rightarrow Z h^{0}$.

Searches for Higgs bosons with masses beyond the expected 
LEP II sensitivity have probably to wait for the LHC 
or perhaps even for a future high 
luminosity high energy linear $e^{+}e^{-}$ collider. 
\begin{figure}[htb]
\begin{center} 
\epsfig{file=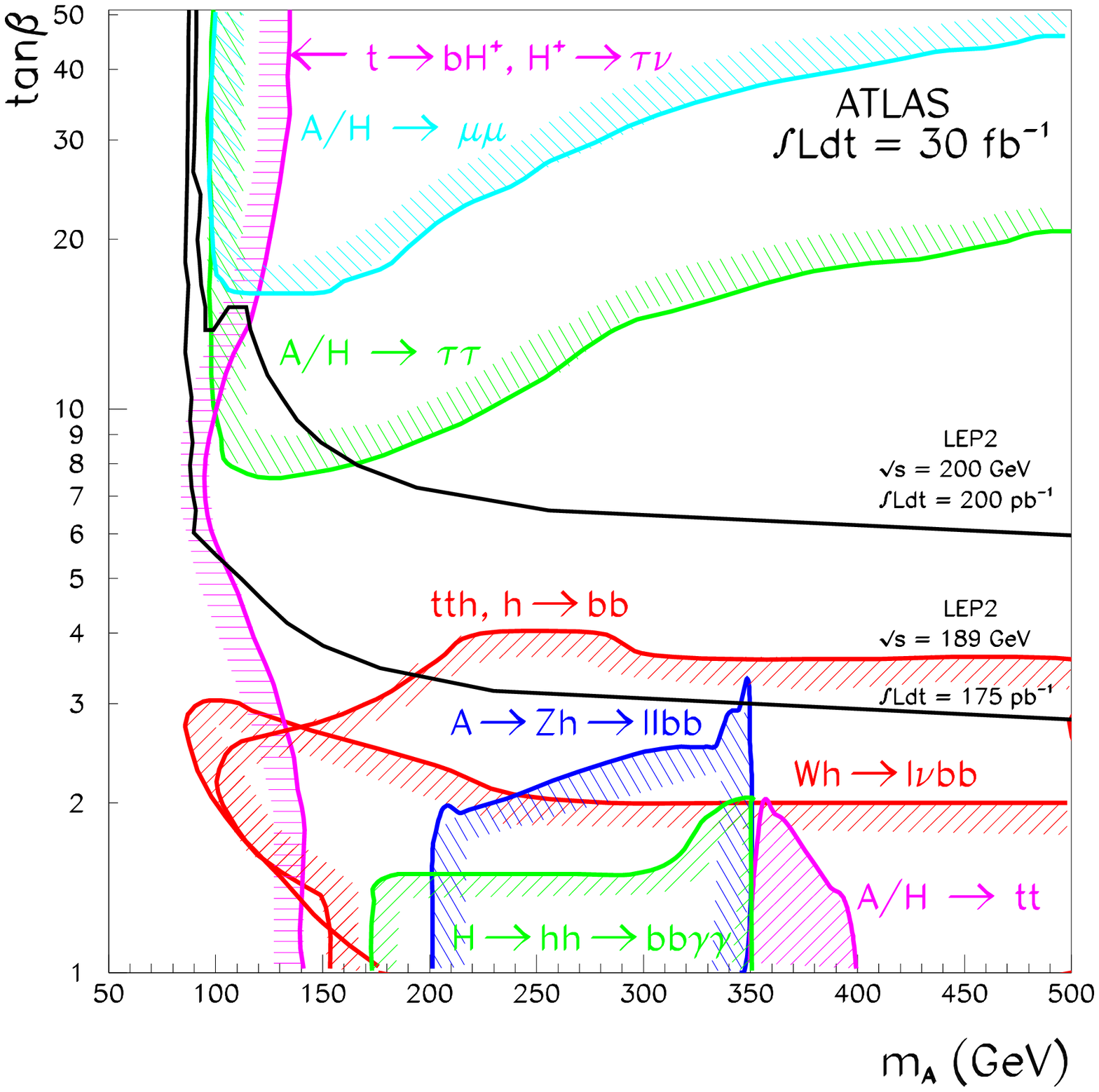,
width=12.cm,height=8.cm}
\epsfig{file=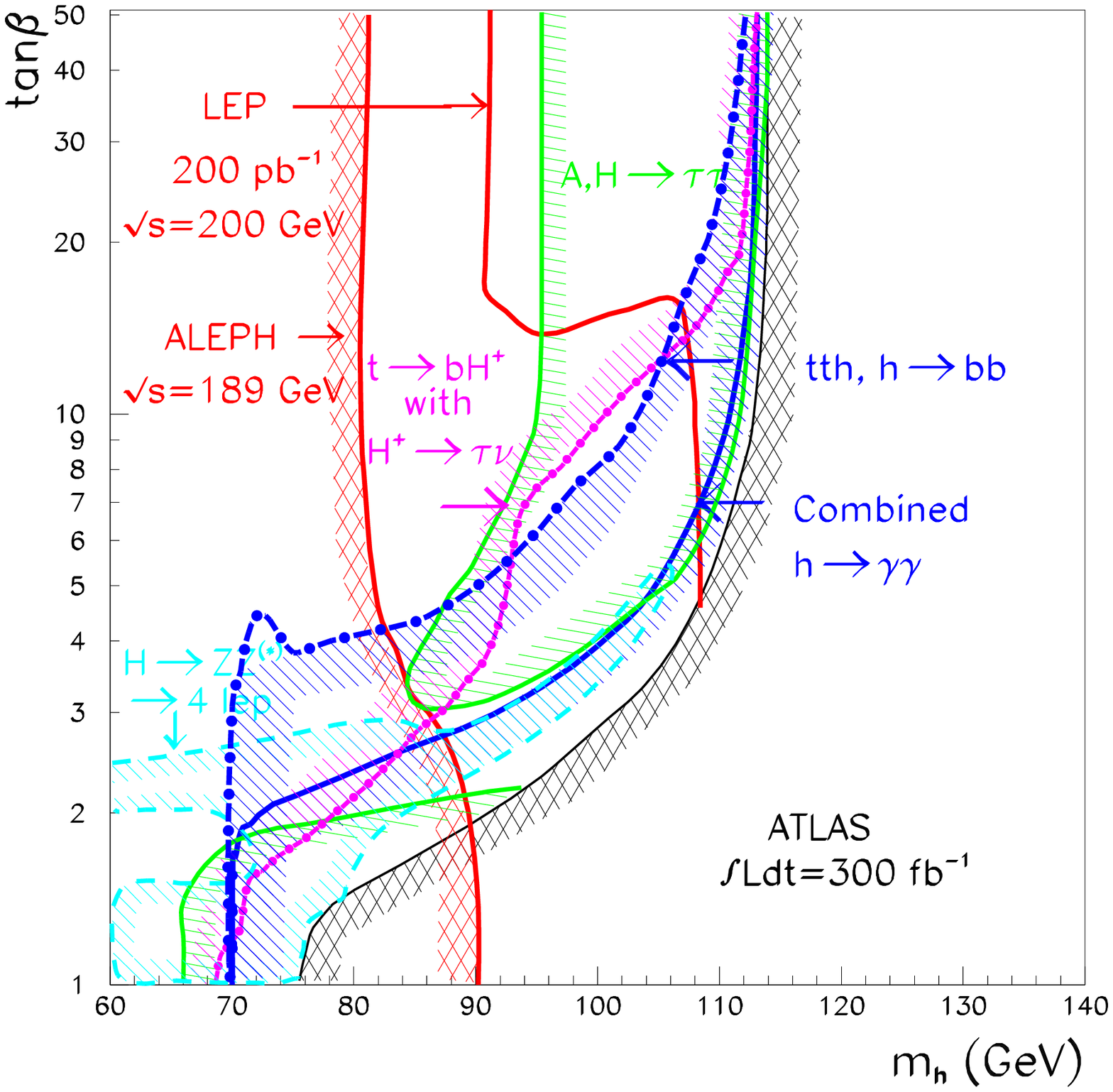,
width=12.cm,height=8.cm}
\caption[fig12]
{ATLAS~\cite{atlas99tdr} 
5 sigma significance contour plot for the different MSSM Higgs 
sector in the $M_{A}$ - $\tan\beta$ plane and for 
30 fb$^{-1}$. The lower plot shows the
ultimate ATLAS 5 sigma discovery sensitivity for 
the MSSM with a luminosity of 300 fb$^{-1}$ 
in the $M_{h}$ - $\tan\beta$ plane. 
Each curve indicates the sensitivity for different 
Higgs search modes.} 
\end{center}
\end{figure}

Todays sensitivity studies from both large LHC experiments,  
ATLAS~\cite{atlastp} and CMS~\cite{cmstp}, 
indicate an excellent sensitivity to  
SM Higgs boson up to masses of about 1 TeV. 
The sensitivity to the Higgs sector of the MSSM scenario 
appears currently to be somehow restricted. 
For the lightest Higgs, with a mass below 120--130 GeV, the only established
signature is the decay $h^{0} \rightarrow \gamma \gamma$. 
For masses of $M_{A}$, greater than 400 GeV 
the $h^{0}$ behaves essentially like the  
SM Higgs rates and should be discovered a few years after the LHC start.
For smaller masses of $M_{A}$, 
the branching ratio $h \rightarrow \gamma \gamma$ becomes 
too small to be observable and 5 standard deviation $h^{0}$ signals 
can only be expected from 
the combination of the $h^{0} \rightarrow \gamma \gamma$ search 
with other $h^{0}$ decay modes, like 
$pp \rightarrow t\bar{t} h^{0} \rightarrow b \bar{b}$,
$h^{0} \rightarrow ZZ^{*}$ and 
$h^{0} \rightarrow WW^{*}$.
The regions where one should see 5 standard deviation MSSM Higgs 
signals and with integrated luminosity of 30 fb$^{-1}$, about three 
years after the LHC start and with a ``final'' luminosity of 
300 fb$^{-1}$ are indicated in Figures 12a and b, respectively.  
The expected sensitivity of the LHC experiments 
to other Higgs bosons of the MSSM are also indicated in Figure 12.

\clearpage 
%\newpage
\subsection {Examples of todays direct SUSY Searches}

Todays searches cover a wide range of MSSM SUSY models, going from 
the most ``conservative'' MSUGRA model to more 
``radical'' assumptions like Gauge mediated models and models 
with R--parity violation. But so far and despite the large variety 
of studied signatures,
no indication for SUSY like particles has been found at LEP II, 
at the TEVATRON or at HERA.

To demonstrate the good experimental sensitivity 
it appears to be useful to discuss the actual outcome 
of a few SUSY searches at LEP II and the TEVATRON. 

The first example is the search for acoplanar lepton pairs events
from OPAL at LEP II~\cite{opalsleptons}. The distribution of the 
lepton energy scaled by the beam energy and the reconstructed 
scattering angle multiplied by the charge of the most energetic lepton 
with respect to the incoming electron direction 
are shown in Figures 13a and b respectively.
The data are in agreement with expectations from 
the dominant $WW$ backgrounds. The possibility 
to discriminate potential signal events from 
the backgrounds using the characteristic charge dependent angular
distribution is nicely seen in Figure 13b.

\begin{figure}[htb]
\begin{center}
\epsfig{file=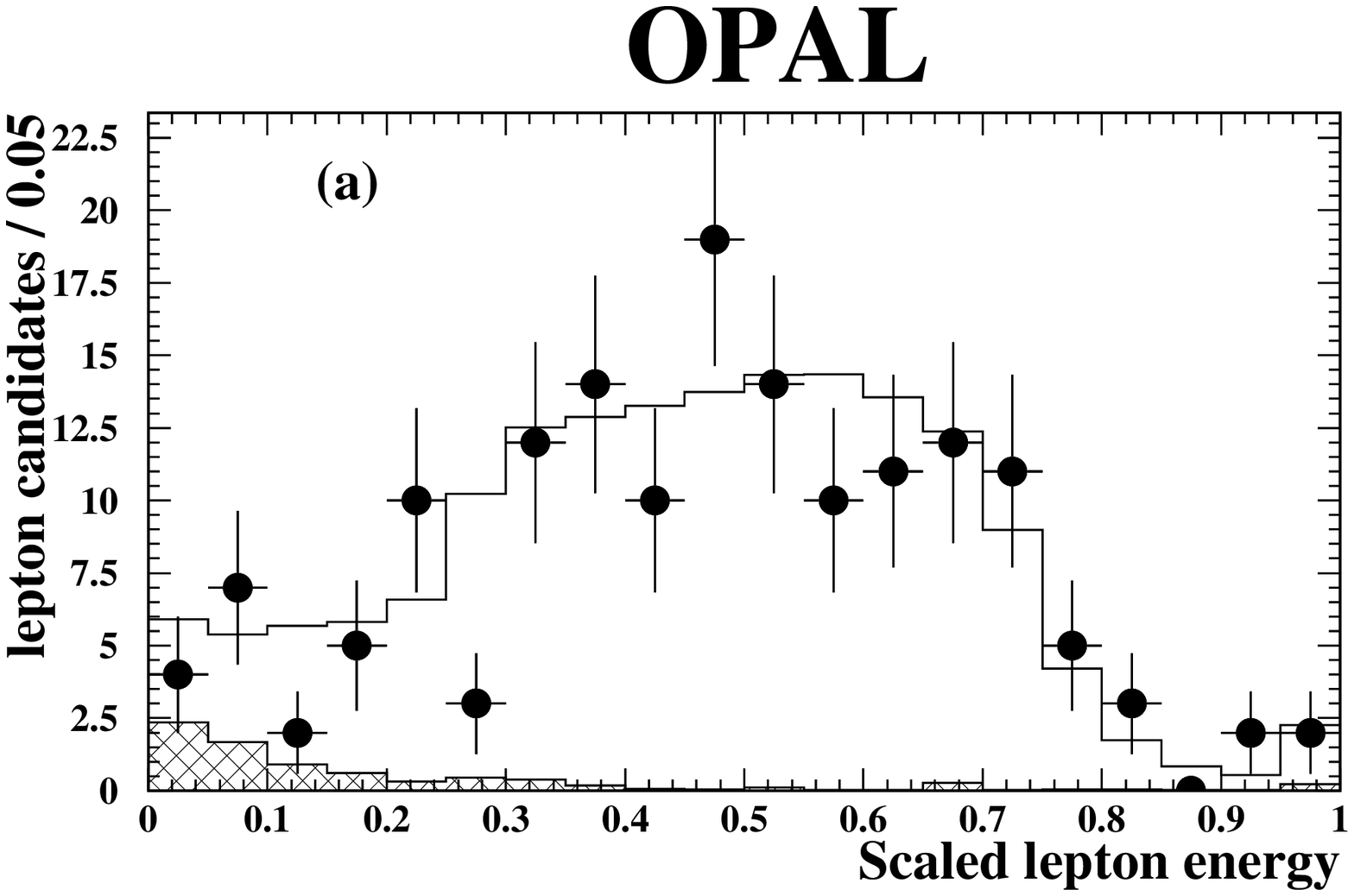,
width=7.cm,height=6.cm}
\epsfig{file=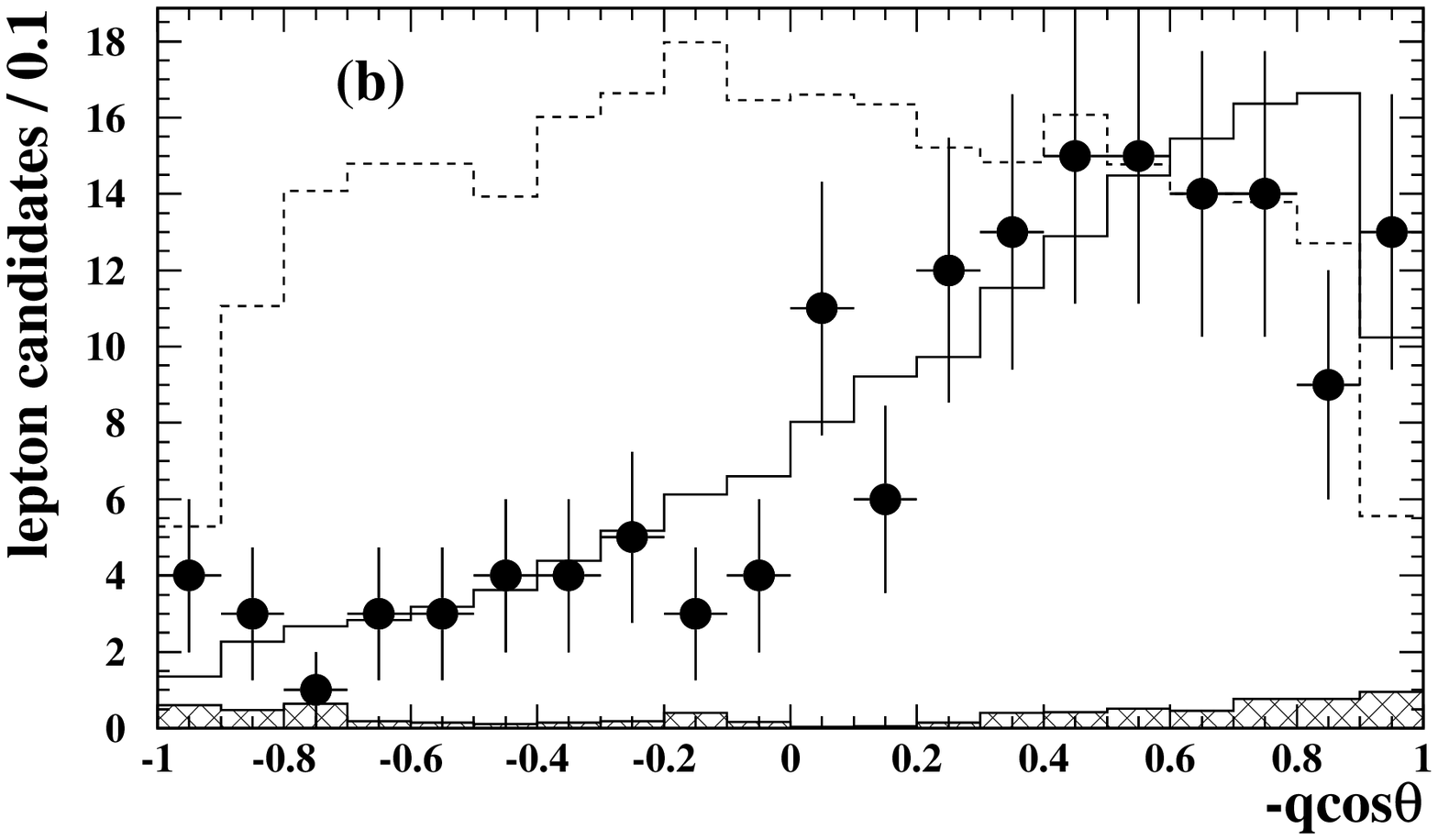,
width=7.cm,height=6.cm}
\caption[fig13]
{Observed and expected distributions of acoplanar lepton pair events 
from OPAL~\cite{opalsleptons}. Shown are the scaled 
lepton energy ($E_{\ell^{\pm}}/E_{Beam}$) (a) and (b) the reconstructed 
scattering angle multiplied with the associated charge 
with respect to the $e^{-}$ beam direction.}  
\end{center}
\end{figure}

The second example are results from L3, summarized in 
Table 1~\cite{l3susy1}.
The observed number of events in the data are compared with expected 
backgrounds and optimized searches for sleptons and charginos,
with small, medium and large mass differences between 
the lightest stable SUSY particle and the studied SUSY particle.
Not even a two sigma excess is seen.
An interpretation of the L3 results, within the MSSM model, is given in 
Figure 14. 
\begin{table}[htb]
  \begin{center}
    \begin{tabular}{|c|c|c|c|c|c|c|}\hline
        \multicolumn{7}{|c|}{$\sqrt{s}=189$ GeV data} \\ \hline
        & \multicolumn{2}{|c|}{Low $\Delta$ M}
        & \multicolumn{2}{|c|}{Medium $\Delta$ M}
        & \multicolumn{2}{|c|}{High $\Delta$ M}\\ \hline
& N$_{Data}$ & N$_{exp}$
& N$_{Data}$ & N$_{exp}$
& N$_{Data}$ & N$_{exp}$ \\ \hline
$\tilde{e}$ & 7   &  6  &  3 & 4.8 & 11 & 12.4 \\ \hline
$\tilde{\mu}$ & 10   &  11.5  &  2 &  1 &  8 & 9.1 \\ \hline
$\tilde{\tau}$ & 4   &   2.7  &  3 &  8.4 &  9 & 11.9 \\ \hline
$\tilde{\chi^{\pm}}$ &  72     &  66.9  & 11 &  10.9 &  67 & 76.7 \\ \hline
$\tilde{\chi^{0}}_{2}$ & 43      &  39.3  & 6 &  7.78 &  3 & 2.45 \\ \hline
\hline
    \end{tabular}
    \caption{Observed and expected (SM) number of events 
for slepton, chargino and neutralino searches with L3 at LEP~\cite{l3susy1}.
\label{tab:table1}}
  \end{center}
\end{table}
\begin{figure}[htb]
\begin{center}
\epsfig{file=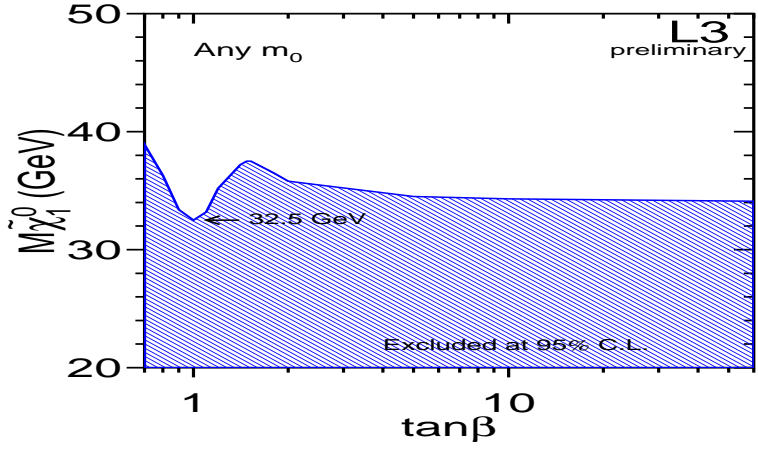,
width=12.cm,height=6.5cm}
\caption[fig14]
{Excluded mass range for $\tilde{\chi^{0}_{1}}$ and $\tan \beta$
obtained from the combined chargino and neutralino searches in L3 
and the data taken at $\sqrt{s} = 189 $ GeV~\cite{l3susy1}. 
}
\end{center}
\end{figure}

The third example is related to a search for trilepton events,  
originating from the reaction 
$q_{i}\bar{q}_{j} \rightarrow \tilde{\chi}^{0}_{2}\tilde{\chi}^{\pm}_{1}$ 
and leptonic decays of the neutralino and chargino. 
Such events can be detected from an analysis of events with 
three isolated high $p_{t}$ leptons and large missing transverse energy. 
The potential of this trilepton signature at hadron colliders 
like the LHC has been described in several phenomenological 
studies~\cite{trileptons}. It was found, that trilepton events 
with jets need to be rejected in order to
distinguish signal events from SM and other SUSY backgrounds.

After the removal of jet events, the only remaining relevant 
background comes from leptonic decays of $WZ$ events.
Potential backgrounds from dilepton events like 
$W^{+}W^{-} \rightarrow \ell^{+} \nu \ell^{-} \bar{\nu}$ and 
hadrons misidentified as electrons or muons 
are usually assumed to be negligible.
Depending on the analyzed SUSY mass range, the background from 
leptonic decays of $WZ$ events, in contrast to a 
potential signal, will show a $Z^{0}$ mass peak in the dilepton 
spectrum. 
The results of a TEVATRON (CDF) trilepton search~\cite{cdftrileptons}
are shown in Table 2.
\begin{table}[htb]
% \vspace{0.4cm}
\begin{center}
\begin{tabular}{|c|c|c|c|}
\hline
Cut           & observed  & SM Background & MSSM MC \\
              & Events    & Expectation & $M(\tilde{\chi}_{1}^{\pm})=
M(\tilde{\chi}_{2}^{0})$ =70 GeV \\
\hline
Dilepton data                & 3270488 &  &   \\
\hline
Trilepton data               &     59  &  &  \\
\hline
Lepton Isolation             &     23  &  &   \\
\hline
$\Delta R_{\ell \ell} > 0.4$ &      9  &  &   \\
\hline
$\Delta \phi_{\ell\ell}<170^{o}$ &  8 & 9.6$\pm$1.5 & 6.2 $\pm$0.6  \\
\hline
$J/\Psi, \Upsilon, Z$ removal &     6 & 6.6$\pm$1.1 & 5.5 $\pm$0.5  \\
\hline
missing $E_{t}(miss) > 15 $  &      0 & 1.0$\pm$0.2 & 4.5 $\pm$0.4  \\
\hline
\end{tabular}
\caption{Results from a recent trilepton 
analysis from CDF with a dataset of $\approx$ 
100 pb$^{-1}$~\cite{cdftrileptons}.
The number of observed events shows good agreement with various SM 
background sources.}
\label{tab:table2}
\end{center}
\end{table}

The last example is related to the famous lonely CDF event, 
which has large missing transverse energy,
2 high $p_{t}$ isolated photons and 2 isolated high $p_{t}$ 
electron candidates~\cite{cdfsusyevent}.
The presence of high $p_{t}$ photons does not match MSUGRA
expectations but might fit 
into so called gauge mediated symmetry breaking models, GMSB~\cite{gmsbth}. 
This event has motivated many additional, so far negative searches.  
Particular sensitive searches at LEP II 
come from the analysis of events with one or more 
energetic photons and nothing else.
Such events can originate essentially only from 
initial state bremsstrahlung in the reaction 
$e^{+}e^{-} \rightarrow Z \gamma \gamma$ and the $Z$ decaying to neutrinos
or from the neutralino pair production and subsequent decays to 
photons and invisible gravitinos. No excess of such events has been 
seen by any of the LEP experiments. These negative results 
exclude essentially the SUSY interpretation of the CDF event. 
Typical results, here from L3, for the recoil mass distribution 
of single and double $\gamma$ events and their interpretation 
in comparison with the area allowed by the CDF event,  
are shown in Figures 15a-c~\cite{l3susy2}.

\begin{figure}[htb]
\begin{center}
\epsfig{file=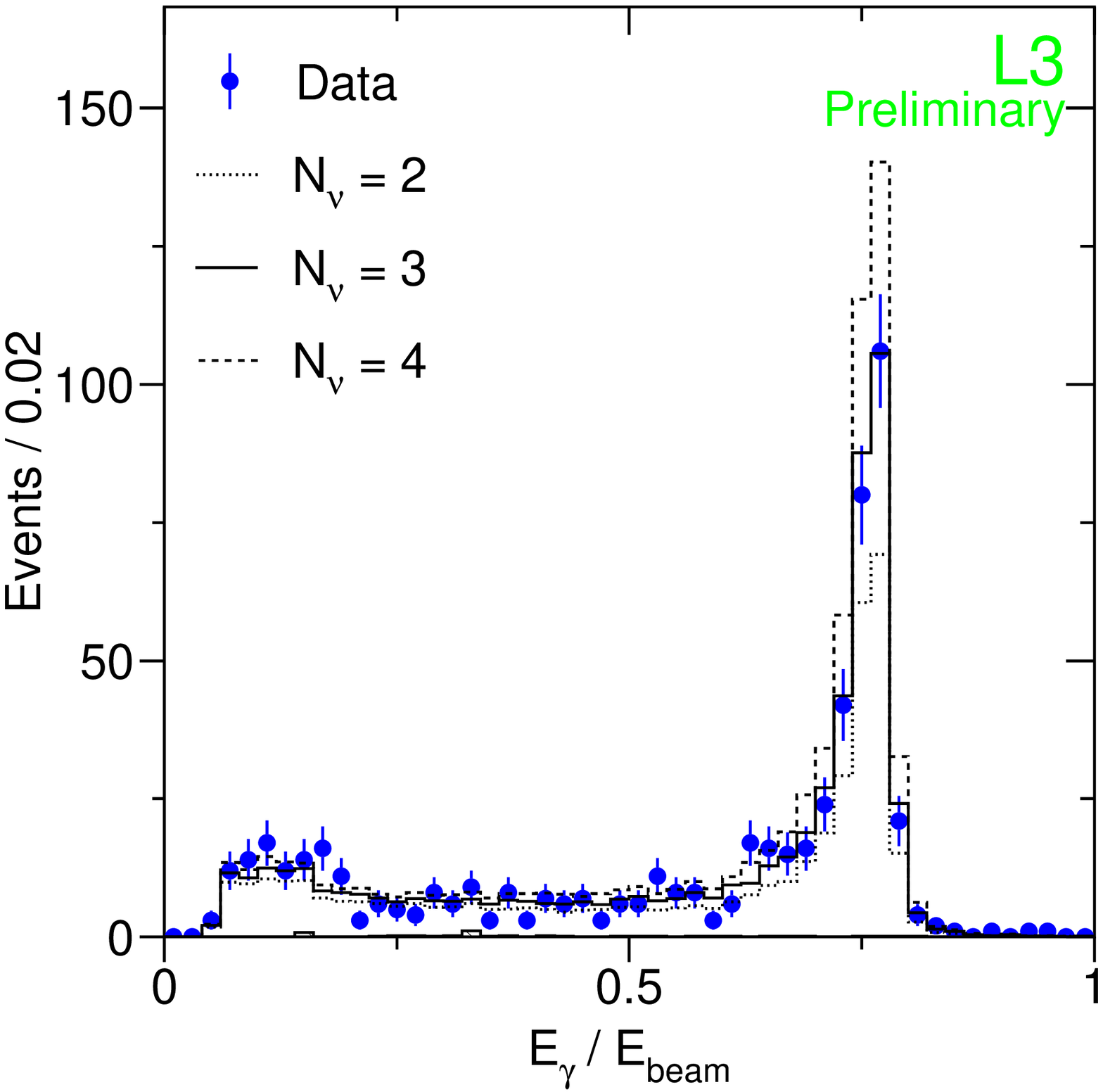,
width=7.cm,height=8.cm}
\epsfig{file=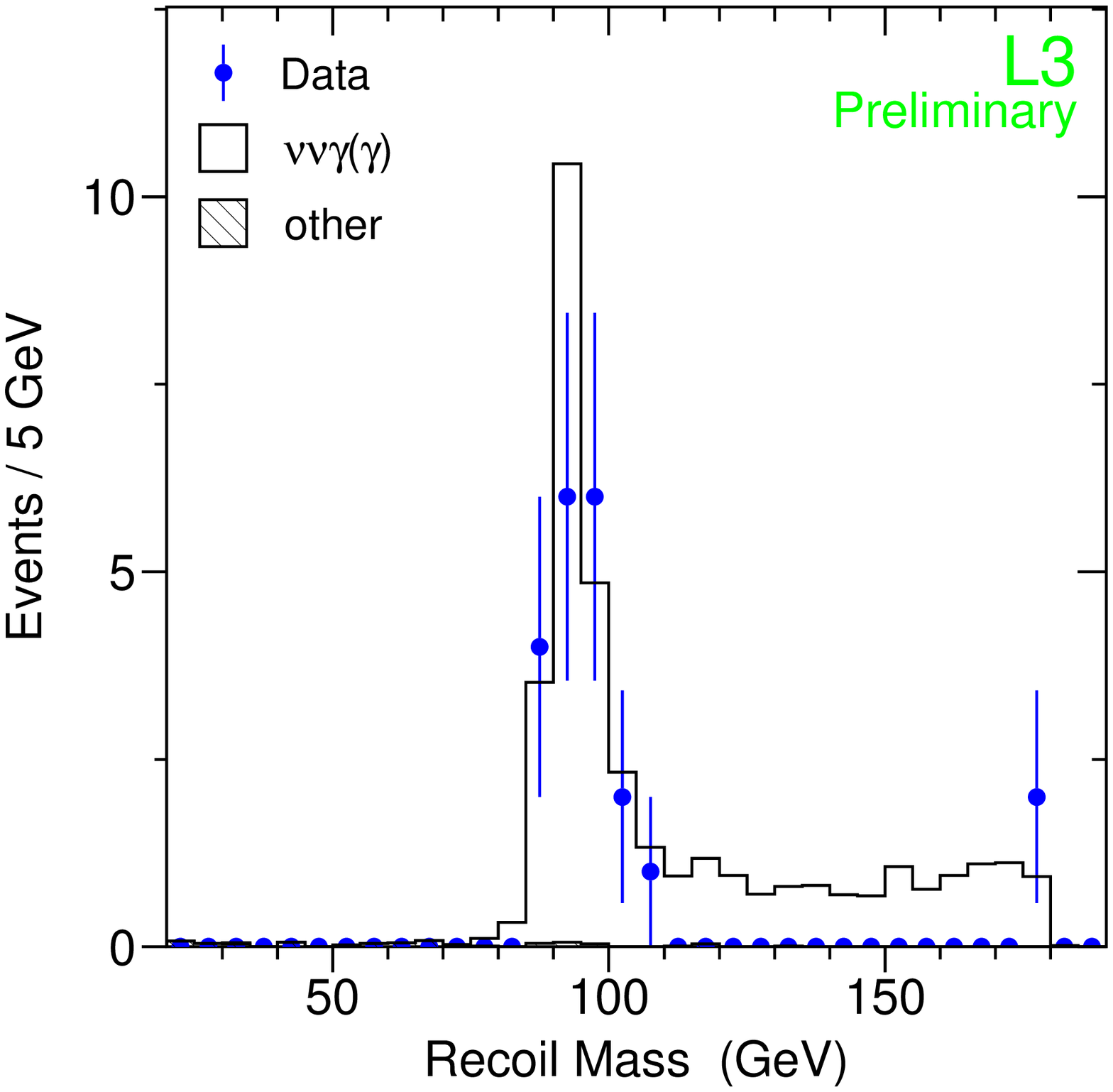,
width=7.cm,height=8.cm}
\epsfig{file=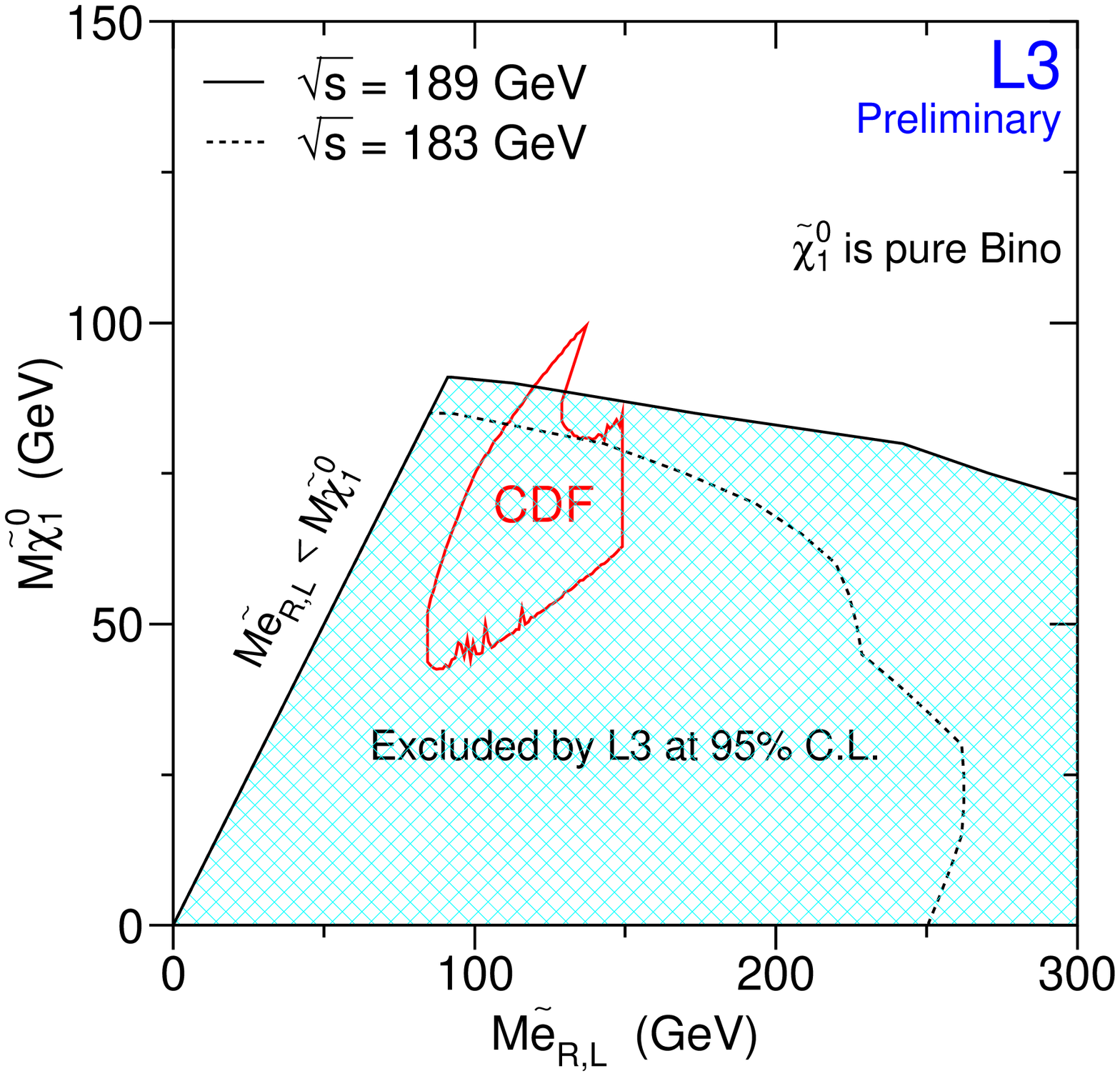,
width=14.cm,height=10.cm}
\caption[fig15]
{The scaled energy and recoil mass distribution of events 
which consist of single or double photons as observed and 
expected by the L3 experiment~\cite{l3susy2}. 
The lower plot shows the interpretation 
of the LEP II data within the assumed model and in comparison with 
the CDF event.
}
\end{center}
\end{figure}
\clearpage
\subsection{Where we might discover SUSY particles} 

The data taking at LEP II will continue during the year 2000, with an 
expected maximal $\sqrt{s}$ of  $\approx 200$ GeV.
In contrast to the MSSM Higgs search, 
where a still a sizeable fraction of the parameter space can be covered, 
the increase in the mass range from $\approx$ 90 GeV to 95 GeV
appears to be small compared to possible 
TeV scale masses of SUSY particles.

The next phase of direct SUSY searches will thus be dominated by 
hadron colliders. The 
improved TEVATRON collider is expected to start data taking 
during the year 2000. The expected yearly luminosity for the so called 
RUN II should reach a few fb$^{-1}$ per experiment. This should allow to 
discover the reaction 
$\tilde{\chi}^{0}_{2}\tilde{\chi}^{\pm}_{1}$ with SUSY masses 
up to 130 GeV. Further improvements could come from the third
phase of the TEVATRON (RUN III),   
which could reach chargino masses up to about 210 GeV
and integrated luminosities of 20--30 fb$^{-1}$~\cite{tev2000susy}. 
The final test of the MSSM version of Supersymmetry
should come from CERN's LHC, currently expected to start operation
during the year 2005. LHC experiments are especially sensitive 
to strongly interacting particles with their huge 
production cross section. 
For example, the pair production cross section of 
squarks and gluinos with a mass of $\approx$ 1 TeV has been estimated 
to be as large as 1 pb resulting in 
10$^{4}$ produced SUSY events for one ``low'' luminosity 
LHC year~\cite{cernth96}. 
Depending on the SUSY model, a large variety of 
massive squark and gluino decay channels and signatures might exist.
A complete search analysis for squarks and gluons 
at the LHC should consider the various 
signatures resulting from the following decay channels. 
\begin{itemize}
\item $\tilde{g} \rightarrow \tilde{q} q$ and perhaps 
$\tilde{g} \rightarrow \tilde{t} t$ 
\item $\tilde{q} \rightarrow \tilde{\chi}^{0}_{1} q$ ~~~or~~~
$\tilde{q} \rightarrow \tilde{\chi}^{0}_{2} q$  ~~~or~~~
$\tilde{q} \rightarrow \tilde{\chi}^{\pm}_{1} q$
\item $\tilde{\chi}^{0}_{2} \rightarrow \tilde{\chi}^{0}_{1} \ell^{+}\ell^{-}$ 
 ~~~or~~~ $\tilde{\chi}^{0}_{2} \rightarrow \tilde{\chi}^{0}_{1} Z^{0} $  
~~~or~~~ $\tilde{\chi}^{0}_{2} \rightarrow \tilde{\chi}^{0}_{1} h^{0}$
\item $\tilde{\chi}^{\pm}_{1} \rightarrow \tilde{\chi}^{0}_{1} \ell^{\pm}\nu$
~~~or~~~$\tilde{\chi}^{\pm}_{1} \rightarrow \tilde{\chi}^{0}_{1} W^{\pm}$.
\end{itemize}  
The various decay channels 
can be separated into at least three distinct event signatures.
\begin{itemize}
\item Multi--jets plus missing transverse energy. These events 
should be ``circular'' in the plane transverse to the beam.
\item Multi--jets plus missing transverse energy plus n(=1,2,3,4) 
isolated high $p_{t}$ leptons. These leptons originate 
from cascade decays of charginos and neutralinos.
\item
Multi--jets plus missing transverse energy plus same charge leptons pairs. 
Such events can be produced in events of the type 
$\tilde{g}\tilde{g} \rightarrow \tilde{u} \bar{u} \tilde{d} \bar{d}$
with subsequent decays of the squarks to 
$\tilde{u} \rightarrow \tilde{\chi}^{+}_{1} d$ 
and $\tilde{d} \rightarrow \tilde{\chi}^{+}_{1} u$ followed by
leptonic chargino decays  
$\tilde{\chi}^{+}_{1} \rightarrow \tilde{\chi}^{0}_{1} \ell^{+} \nu$.  
\end{itemize}
It is easy to imagine that the observation and detailed analysis
of the different types of squark and gluino signatures might allow 
to measure some of the many MSSM parameters.

The above signatures have already been investigated with the data from 
the TEVATRON RUN I. The negative searches gave  
mass limits for squarks and gluinos 
up to $\approx 200$ GeV. The estimated 
5--sigma sensitivity for RUN II and RUN III 
reaches values as high as 350--400 GeV.
More details about the considered signal and backgrounds 
can be found from the TeV2000 studies~\cite{tev2000susy} and 
the ongoing TEVATRON RUN II workshop~\cite{newtev2000susy}.
\begin{figure}[htb]
\begin{center}
\includegraphics*[scale=0.6,bb=40 200 600 700 ]
{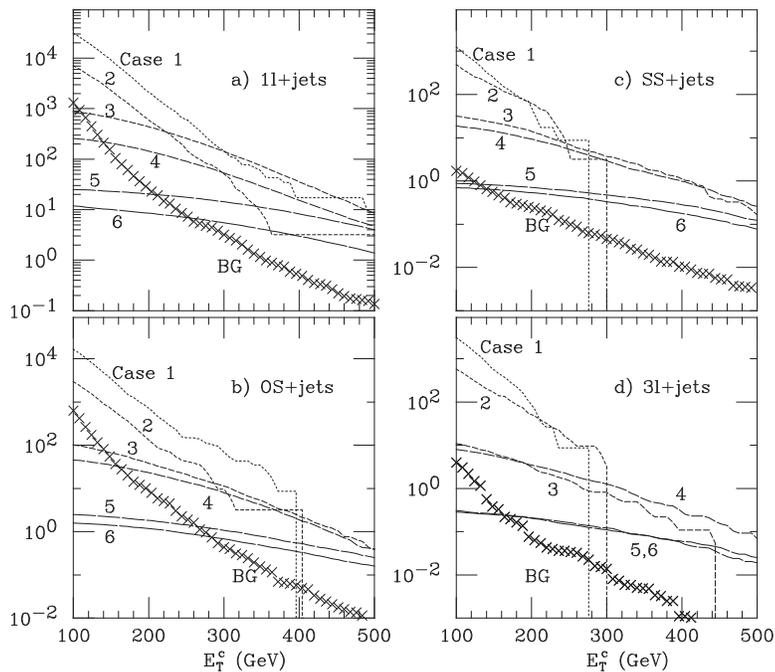}
\caption[fig16]
{Expected $E_{t}^{c}$ distributions for SUSY signal 
and background processes at the LHC and realistic experimental cuts
for $\tan \beta = 2$ and $\mu < 0$~\cite{baer95}.
The different cases are for: 
(1) $m_{\tilde{g}}$= 290 GeV and $m_{\tilde{q}}$= 270 GeV;  
(2) $m_{\tilde{g}}$= 310 GeV and $m_{\tilde{q}}$= 460 GeV;  
(3) $m_{\tilde{g}}$= 770 GeV and $m_{\tilde{q}}$= 720 GeV;
(4) $m_{\tilde{g}}$= 830 GeV and $m_{\tilde{q}}$= 1350 GeV;  
(5) $m_{\tilde{g}}$= 1400 GeV and $m_{\tilde{q}}$= 1300 GeV;
(6) $m_{\tilde{g}}$= 1300 GeV and $m_{\tilde{q}}$= 2200 GeV.}
\end{center}
\end{figure}

A simplified search strategy for squarks and gluinos 
at the LHC would study 
jet events with large visible transverse mass and some missing 
transverse energy. Such events can then be classified according 
to the number of isolated high $p_{t}$ leptons.
Once an excess above 
SM backgrounds is observed for any possible combination of the 
transverse energy spectra, one would try to explain the 
observed types of exotic events and their cross section(s) for
different SUSY $\tilde{g}, \tilde{q}$ masses and decay modes and models.
An interesting approach to such a multi--parameter analysis uses some 
simplified selection variables. For example one could use
the number of observed jets and leptons and their transverse energy, their
mass and the missing transverse energy to separate signal and 
backgrounds. Such an approach has been used to perform a ``complete''
systematic study of $\tilde{g}$ and $\tilde{q}$ decays~\cite{baer95}. 
The proposed variable $E_{t}^{c}$ is the value of the smallest of 
$E_{t}$(miss), $E_{t}$(jet1), $E_{t}$(jet2). The events are further 
separated into the number of isolated leptons. Events with lepton pairs 
are divided into same sign (charge) pairs (SS) and opposite 
charged pairs (OS). 
Signal and background distributions for various squark and gluinos 
masses, obtained with such an approach are shown in Figure 16.  
According to this classification the number of expected 
signal events can be compared with the various SM background processes.
The largest and most difficult backgrounds originate mainly from 
$W+$jet(s), $Z+$jet(s) and $t\bar{t}$ events. 
Using this approach, very encouraging
signal to background ratios, combined with quite large signal cross sections 
are obtainable for a large range of squark and gluino masses.
The simulation results of such studies indicate, as shown in Figure 17,
that the LHC experiments are sensitive to 
squark and gluinos masses up to masses of about 2 TeV, 
$m_{\tilde {g}} = 3 \cdot m_{1/2}$, and a luminosity of 100 fb$^{-1}$. 

\begin{figure}[htb]
\begin{center}
\includegraphics*[scale=0.6,bb=40 320 600 800 ]
%{/home/xv/dittmar/zuoz/baer1.ps}
{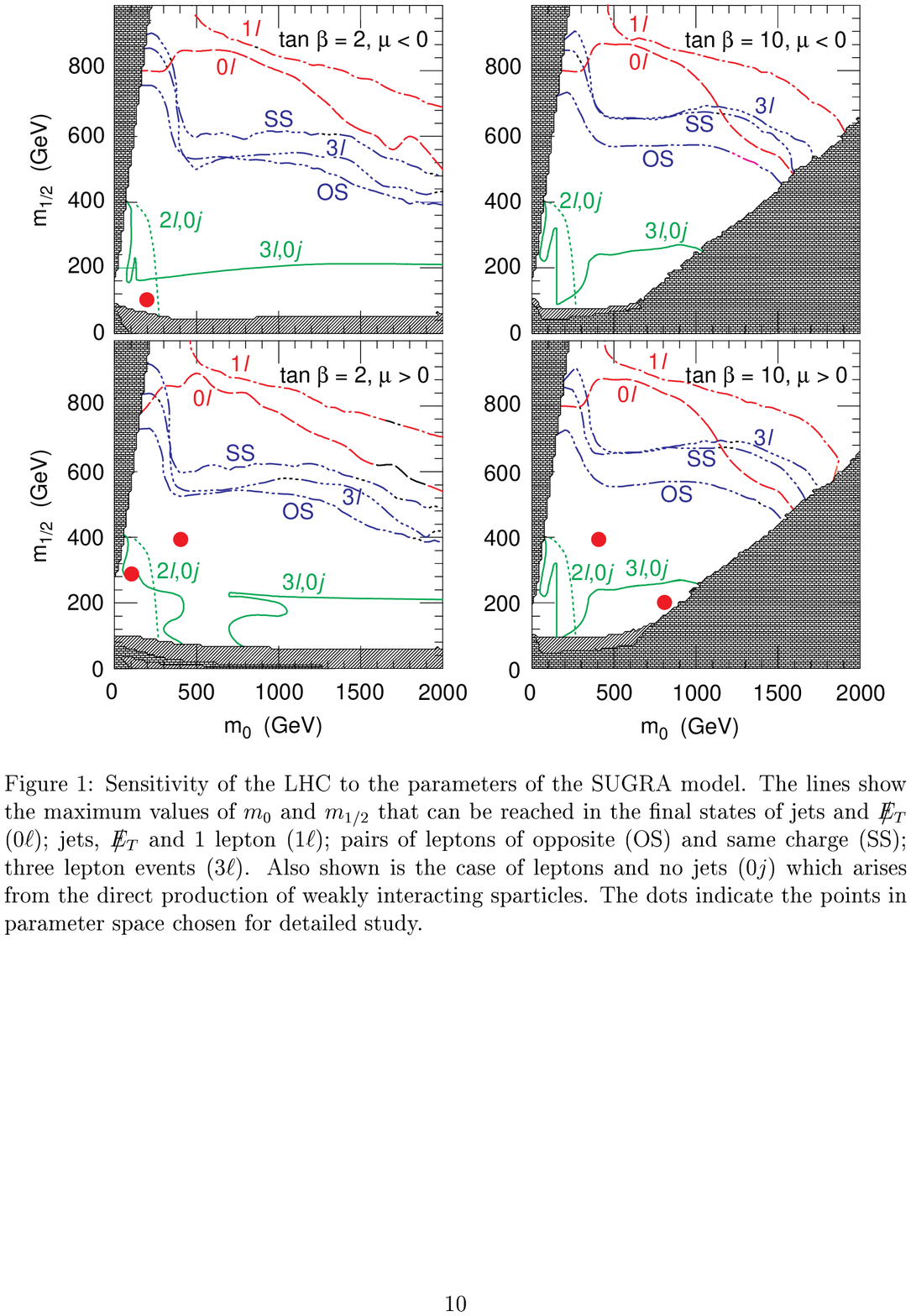}
\caption[fig31]
{Expected sensitivity for various SUSY particles and signatures 
in the $m_{0} - m_{1/2}$ plane using an integrated luminosity of 
10 fb$^{-1}$ at the LHC~\cite{baer95}. The different curves indicate 
the expected sensitivity for SUSY events with n leptons ($\ell$) 
and for events with lepton pairs with same charge (SS) and 
opposite charge (OS).}  
\end{center}
\end{figure}

Figure 17 indicates further, that detailed studies of branching ratios 
are possible up to squark or gluino masses of about 1.5 TeV, where 
significant signals can be observed with many different channels.   
Another consequence of the expected large signal cross sections
is the possibility that the ``first day'' LHC luminosity 
$\approx$ 100 pb$^{-1}$ should be sufficient to discover 
squarks and gluinos up to masses of about 600--700 GeV, 
well beyond even the most optimistic 
TEVATRON Run III mass range.

\subsection{SUSY discovered, what can be studied at the LHC?}

Being convinced of the LHC SUSY discovery potential, 
one certainly wants to know if 
``the discovery'' is consistent with Supersymmetry and
if some of the many SUSY parameters can be measured.   
To answer such a question one should try find many
SUSY particles and measure their decay patterns as accurately as 
possible. For example one finds that the
production and decay of $\tilde{\chi}^{0}_{2}\tilde{\chi}^{\pm}_{1}$
provides good rates for a trilepton signature if the chargino and neutralino 
masses are below 200 GeV. The observation of such events
should allow to measure accurately the dilepton mass 
distribution which is sensitive
to the mass difference between the two neutralinos.
Depending on the used MSUGRA parameters one finds that the 
$\tilde{\chi}^{0}_{2}$ can have two or three body decays. 
The relative $p_{t}$ spectra of the two leptons
can be used to distinguish the two possibilities.

In contrast to the rate limitations of weakly produced SUSY 
particles at the LHC, detailed 
studies of the clean squark and gluino events are expected 
to reveal much more information.  
One finds that the large rate for many 
distinct event channels allows to measure masses and mass 
ratios for several SUSY particles, which are possibly 
being produced in cascade decays of squarks and gluons. 
Many of these ideas have been discussed at a 1996 
CERN Workshop~\cite{cernth96}.
Especially interesting appears to be the idea that 
the $h^{0}$ might be produced and detected in the decay chain 
$\tilde{\chi}^{0}_{2} \rightarrow \tilde{\chi}^{0}_{1} h^{0} $ and 
$h^{0} \rightarrow b \bar{b}$. The simulated mass distribution 
for $b \bar{b}$ jets 
in events with large missing transverse energy is shown 
in Figure 18. Clear Higgs mass peaks above background are found 
for various choices of $\tan \beta$ and $m_{0}, m_{1/2}$.  

\begin{figure}[htb]
\begin{center}
\includegraphics*[scale=1.5,bb=100 300 550 750 height=23. cm,width=18.cm]
{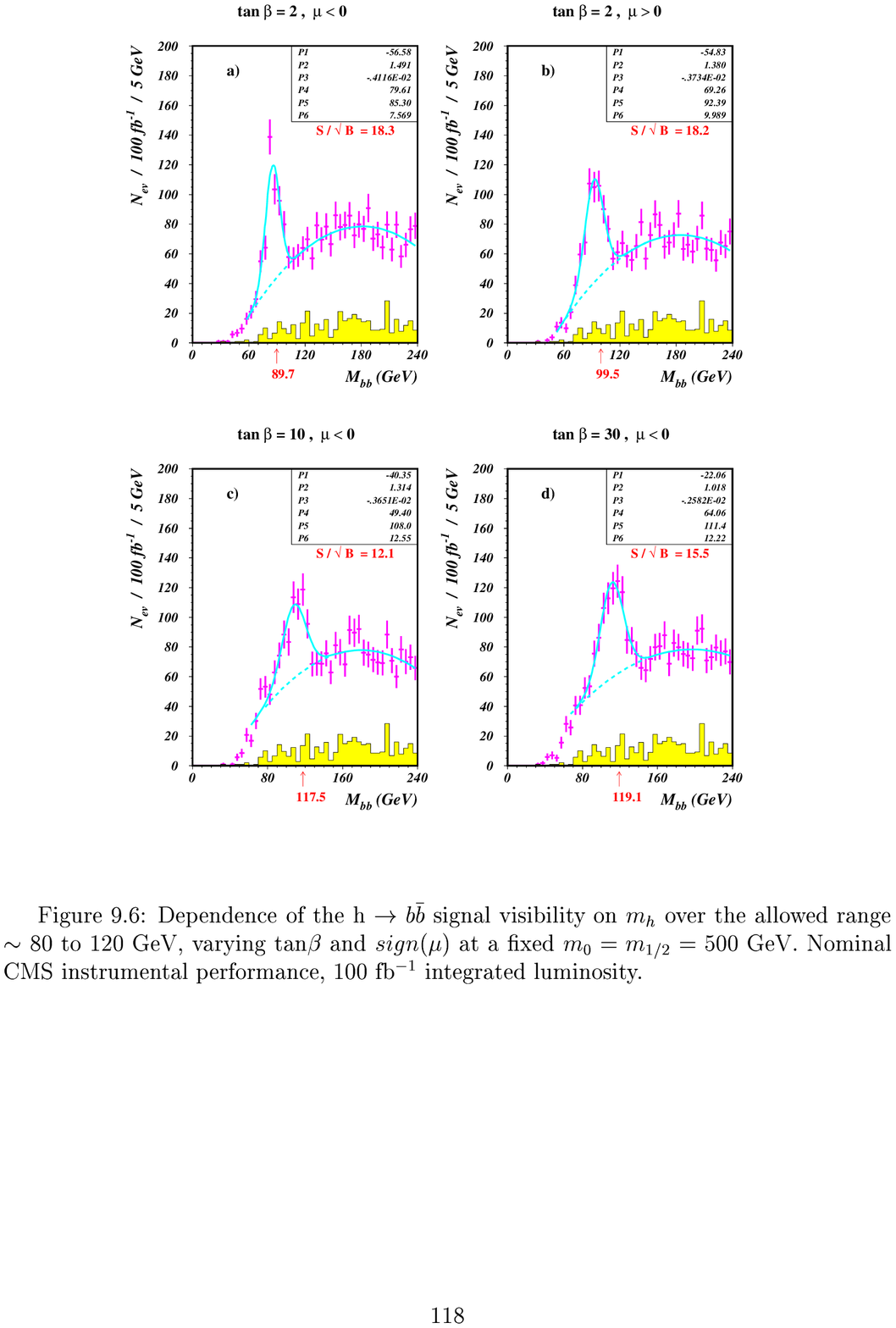}
\caption[fig18]
{Possible inclusive Higgs signals in squark and gluino events,
reconstructed from the invariant mass of $h \rightarrow b \bar{b}$ 
with CMS and a luminosity of  
L=100 fb$^{-1}$~\cite{cms98006}.}
\end{center}
\end{figure}
 
\clearpage
\section{Summary}
Searches for 
Supersymmetry at the highest energy 
particle colliders can be divided into the  
search for the MSSM Higgs sector and for the direct search for SUSY
particles. So far no signs of neither a Higgs boson nor of any 
SUSY particles have been found. 

The expected energy increase of the LEP II collider during the year 2000
might be just right to 
detect a Higgs boson up to a mass of about 105 GeV.
In contrast, it appears that 
the LEP II experiments have reached 
almost the kinematical limit for the direct detection 
of supersymmetric particles as 
only marginal improvements, about 5\%,  
can be expected from the future LEP II running.

The future high luminosity running of the TEVATRON might 
improve the existing sensitivity for 
SUSY particles by a factor of about 1.5--2 compared to todays
mass limits. Consequently charginos might be seen up 
to masses of 200 GeV and  
squarks and gluinos up to masses of about 400 GeV.
This reach should be compared with the expectations from 
the future LHC experiments. ATLAS and CMS studies indicate  
a good sensitivity up to masses of about 2 TeV.
In addition, the detectable LHC squark and gluino cross sections, even for 
moderate masses well above any possible TEVATRON limit, are huge and 
LHC SUSY discoveries might be possible even with a luminosity of 
a few 100 pb$^{-1}$ only, obtainable almost 
immediately at the LHC switch on. 

To finish this report on ``Searches for Supersymmetry'' we would like 
quote a few authorities:
\newline
\vspace{0.05cm}

{\it ``Experiments within the next 5--10 years will enable us to decide 
whether Supersymmetry, as a solution to the naturalness problem of the 
weak interaction is a myth or reality''} H. P. Nilles 1984~\cite{mssm84}  
\newline
\vspace{0.05cm}

{\it ``One shouldn't give up yet'' .... ``perhaps a correct statement is:
it will always take 5-10 years to discover SUSY''} 
H. P. Nilles 1998~\cite{nilles98}
\newline
\vspace{0.05cm}

{\it ``Superstring, Supersymmetry, Superstition''} Unknown
\newline
\vspace{0.05cm}

{\it ``New truth of science begins as heresy, advances to orthodoxy and ends 
as superstition''} T. H. Huxley (1825--1895).
\newline
\vspace{0.05cm}

{\bf \large Acknowledgments}
{\small I would like to thank the organizers of
the Colloque Cosmologie for the possibility to review
``Supersymmetry searches with existing and 
future collider experiments''.  This invitation gave me the opportunity 
to listen to many interesting presentations about 
the status and open questions of cosmology.   
}
 
\end{document}